\documentclass[nofootinbib,aps,prl,a4paper,twocolumn,english,
superscriptaddress,longbibliography,reprint]{revtex4-2}
\usepackage{amsmath,amssymb,amsfonts,mathrsfs, amsbsy}
\usepackage{graphicx}
\usepackage[bookmarks=true, colorlinks=true, linkcolor=blue, urlcolor=blue, citecolor=blue, bookmarks=true, hyperindex=true]{hyperref}
\usepackage[normalem]{ulem}
\usepackage{physics}
\usepackage{comment}
\usepackage{subcaption}
\usepackage[dvipsnames]{xcolor}
\usepackage[utf8]{inputenc}

\newcommand{\be}{\begin{equation}}
\newcommand{\ee}{\end{equation}}

\newcommand{\beq}{\begin{eqnarray}}
\newcommand{\eeq}{\end{eqnarray}}

\def\H1{\widehat{H}_1}

\begin{document}

\title{Exactly solvable Russian-doll model: Renormalization-group cycles meet fractality}

\author{Ilya Liubimov}
\affiliation{Landau Phystech School of Physics and Research, Moscow Institute of Physics and Technology, 141700, Dolgoprudny, Russia}
\affiliation{
Skolkovo Institute of Science and Technology, 121205, Moscow, Russia}
\affiliation{
Phystech School of Applied Mathematics and Computer Science, Moscow Institute of Physics and Technology, 141700, Dolgoprudny, Russia
}
\author{Alexander Gorsky}
\affiliation{
Phystech School of Applied Mathematics and Computer Science, Moscow Institute of Physics and Technology, 141700, Dolgoprudny, Russia
}
\affiliation{Institute for Information Transmission Problems RAS, 127051 Moscow, Russia}  
\affiliation{Laboratory of Complex Networks, Center for Neurophysics and Neuromorphic Technologies
}



\begin{abstract}

We consider the Bethe ansatz integrable Russian Doll (RD) model of superconductivity with time-reversal symmetry breaking, which exhibits a cyclic renormalization group. By obtaining an exact solution for the renormalization group flows, we investigate the phase structure in the one-pair sector, which includes localized, fractal, and delocalized phases. We show that the quantum number $Q$, arising from the Bethe ansatz equations, counts the number of cycles and parametrizes the towers of states. Using the action of the renormalization group on the eigenstates, we demonstrate that $Q$ serves as an order parameter, providing a new mechanism
for the formation of the fractal phase in the deterministic systems and
an example of the interplay between fractality and cyclic RG.

\end{abstract}

\maketitle

The fractality of wave functions at the  critical disorder  has been widely discussed in the context of the Anderson transition \cite{evers2008anderson}. More recently, several disordered matrix models with extended fractal phase
have been found \cite{kravtsov2015random, khaymovich2020fragile, monthus2017multifractality, biroli2021levy, Bogomolny2018eigenvalue, faoro2019non, motamarri2022localization,cugliandolo2024multifractal, sharipov2026hilbert, sarkar2023tuning}. 
There are also examples of deterministic models with fractal phase \cite{das2025emergent}. In this context, the emergence of the Russian Doll model (RDM) is of particular interest. RDM was introduced
in the context of superconductivity in finite-dimensional systems \cite{ Leclair2004russian,leclair2004log, Leclair2003russian,  Dunning2004integrability}. It is a generalization of the Richardson model \cite{Richardson1963restricted, Richardson1964exact,ortiz2005bcs, dukelsky2004colloquium} 
{that includes an extra coupling parameter, leading to broken time-reversal symmetry (TRS). It is also} the particular limit of the model of anyon pairing \cite{dunning2010exact}.
The Richardson, RD model and the model of anyon pairing  are Bethe Ansatz (BA) integrable;
the BA equations in the Richardson model coincide with those for the twisted Gaudin model \cite{dukelsky2004colloquium},  the BA equation in RDM coincide with those for the non-homogeneous twisted XXX spin chain  \cite{Dunning2004integrability} and the XXZ spin chain for anyon pairing model \cite{dunning2010exact}.

In \cite{gorsky2025theta} it was shown that the deterministic RD model
exhibits localized, fractal, and delocalized phases in the Hilbert space of the interacting fermionic system. The probability distribution in the one pair sector has the Breit-Wigner form, with the width determined by the TRS-breaking parameter \cite{gorsky2025theta}. Thus, the model provides an example of a system in which integrability coexists with fractality. Moreover, it was demonstrated in \cite{gorsky2025theta} that the mode number $Q$ is related to the phases of the model.

The RDM has another interesting property; it is one of the simplest examples of a system with the cyclic renormalization group (RG) \cite{Bulycheva2014spectrum, braaten2006universality}. The very precise example of cyclic RG has been found in the one-pair sector 
of anyon pairing model  \cite{glazek2002limit, dunning2010exact}. In RDM the cyclic RG was formulated  
in \cite{Leclair2004russian} and a tower of states with Efimov scaling was identified in both single-pair and many-body cases \cite{anfossi2005elementary}. 
The cyclic RG for the disordered RDM model is refined \cite{motamarri2024refined} and the period of the cyclic RG becomes energy dependent.

The RD model also appears to be closely related to the specific
strong coupling point $\frac{1}{g^2}=0$ in $N_f=2N_C$ $\cal{N}$=2 
{Super Quantum Chromodynamics} (SQCD) in the $\Omega$-background in the NS limit \cite{nekrasov2009supersymmetric}. The single Bethe root sector corresponds to the single vortex string defect, and the BA equation in RDM 
defines the vacua in the 2d vortex worldsheet theory \cite{nekrasov2010quantization, dorey2011quantization}. 
The parameter $\theta$ in RDM was identified in  \cite{gorsky2025theta} 
with the shifted conventional $\theta$-term in 4d SQCD  while the parameter $Q$ corresponds to the electric flux on the vortex string worldsheet.

In this Letter, we obtain an exact solution for the flow of the coupling constants over the full range of parameters. It turns out that the quantum number $Q$, which arises from taking the logarithm of the Bethe ansatz equations, determines both the number of renormalization group cycles and the tower of states in the fractal phase. Moreover, we investigate the relation between fractality and the action of the renormalization group on eigenstates, demonstrating that $Q(\gamma,\theta)$ serves as a phase identifier, where $\theta$ quantifies the strength of TRS breaking and $\gamma$ parametrizes the coupling in the model. This provides an example in which cyclic RG is combined with integrability to describe fractal states.

\textit{Model}. — We consider the Hamiltonian defined as:
\begin{equation}\label{Hamiltonian}
    H = \sum_{n = 1}^{N}(\varepsilon_n - x)b_n^{\dagger}b_n - \sum_{n \neq m }(x+i \mathrm{sign}(n-m)y) b_n^{\dagger}b_m 
\end{equation}
where $b_n^{\dagger} = c_{n+}^{\dagger}c_{n-}^{\dagger}$, $b_n = c_{n-}c_{n+}$ are creation/annihilation operators of Cooper pairs, while $c_{n \pm}^{\dagger},c_{n \pm}$ are creation/annihilation operators of fermions in time-reversal states $\pm$. 
{The real parameter $x$ plays the role of a BCS-like constant, while $y$ is a real TRS-breaking parameter \cite{Leclair2004russian}. $N$ is the number of doubly degenerate single particle energy levels with energies $\varepsilon_i$.} We will use parameterization $re^{i\theta} = x+iy$ and focus on the one-pair case, corresponding to the single Cooper pair with a matrix Hamiltonian, which was considered in the context of excitations over the Fermi sphere in \cite{anfossi2005elementary} :
\begin{equation}
    H_{nm} = \delta_{nm} (\varepsilon_m - x) - r e^{i\theta \mathrm{sign}(n-m)}
\end{equation}
where the spectrum is defined via the logarithm of the Bethe equation: 
\begin{equation}\label{BAeq}
     \sum_{l = 1}^N \arctan{\frac{y}{(E - \varepsilon_l)}} = -\theta + \pi Q 
\end{equation}
{where $E$ is the one-pair energy, $Q$ is an integer. This equation has $N$ independent solutions. Depending on the parameter regime, however, different solutions may have either the same or different values of $Q$.}

\textit{Eigenstates}. — The exact solution yields an expression for 
{the components of} the eigenstates for $l > 1$:
\begin{equation}\label{exact_eig}
    \psi_l = \frac{1}{\sqrt{\sum_{k} \frac{1}{\rho^2_k}}}\frac{e^{-i\varphi_l - 2i\sum_{j = 2}^{l-1}\varphi_j -i\varphi_1 }}{\sqrt{(E -\varepsilon_i)^2 + y^2 }} 
\end{equation}
with parametrization $\rho_i = \sqrt{(E -\varepsilon_i)^2 + y^2 }$, $\varphi_i = \arctan{\frac{y}{E - \varepsilon_i}}$. The eigenstates exhibit a Breit-Wigner form with a width parameter $\Gamma = y$. This ansatz was previously proposed to describe fractal states in \cite{monthus2017multifractality,Bogomolny2018eigenvalue}. However, in their approach, it was obtained as an approximation within perturbation theory, whereas in our case, it is exact. In \cite{Khaymovich2023fractalset} it was shown that the fractal dimension of an eigenstate depends solely on the fractal dimension of the set of its diagonal elements rescaled to the interval [0,1]. In the limit $\theta\to 0$, which corresponds to the Richardson model, $\Gamma$ vanishes. 
For $\varepsilon_i = 0$, the spectrum is indexed by the integer $-N/2 +1 \leq Q\leq N/2 + 1/2$:
\begin{equation}\label{delocalizedtower}
    E_Q  = y\cot \frac{\pi Q-\theta}{N}
\end{equation}
The eigenstates have the form of plane waves with momentum $p = 2\phi_Q = 2\frac{\pi Q - \theta}{N}$
\begin{equation}\label{eq:plain_waves}
\psi_l = \frac{e^{-2i\phi_Q l}}{\sqrt{N}} 
\end{equation}

\textit{Fractality of the RDM model}. — We now discuss the presence of fractality for the deterministic RDM. Recall the definition of the fractal dimension $D_q$ for an eigenstate $\psi_n(i)$:
\begin{equation}\label{fractal_def}
    I_q = \sum_{i} |\psi_i|^{2q} \sim N^{D_q (1-q)}
\end{equation}
Using an explicit form of the eigenstates one can calculate $I_q$. The diagonal elements are taken to be equidistant, given by $\varepsilon_i = \delta(i - \frac{N}{2})$ with the bandwidth $\omega = N\delta$, then the characteristic number of sites $M = {\Gamma}/\delta$ for ${N\Gamma}/\omega \gg 1$. In the following, we will assume the condition $N \gg 1$ for all calculations.
Defining $D_q$ as $\frac{1}{1-q}\frac{\ln{I_q}}{\ln{N}}$ in the thermodynamic limit $N\to\infty$, we can find the fractal dimension with the first corrections \cite{gorsky2025theta, suppl} for $1 \ll y\ll N$:
\begin{equation}\label{D_q_y}
    D_q =  \frac{\ln{y/\delta}}{\ln{N}} + O(\frac{1}{\ln{N}})
\end{equation}
It is convenient to change the parameterization and introduce the parameter $\gamma$: $\gamma = -\frac{\ln {r/\delta}}{\ln{N}}$. This parameterization is widely used for matrix Hamiltonians, for example, \cite{kravtsov2015random}. However, we will take a special form of the large $N$ limit, details are discussed in the Supplement \cite{suppl}. Then $\ln{\frac{y}{\delta}}/\ln{N} = - \gamma + \ln{\sin{\theta}}/\ln{N}$. For large $N$ we obtain the following phase picture:
\begin{itemize}
    \item Localized phase: $\gamma > 0$, $\frac{\Gamma}{\delta} \ll 1$, eigenstate is localized on one  site, $D_q = 0$
    \item Fractal phase: $\gamma \in (-1, 0)$, $N \gg \frac{\Gamma}{\delta} \sim N^{-\gamma} \gg 1$, $D_q = -\gamma$
    \item Delocalized phase: $\gamma < -1$, $\frac{\Gamma}{\delta} \gg N$, $D_q = 1$
\end{itemize}
However, to use this definition of phases in terms of $\gamma$, one needs the condition for the term involving $\theta$ : $\ln{\sin{\theta}}/\ln{N}\ll 1$, i.e., $\theta$ should not tend to the lines of $\theta = 0$ and $\theta = \pi$, where RDM reduces to the Richardson limit.
\begin{figure*}[t]
\begin{minipage}[h]{0.325\linewidth}
\center{\includegraphics[width=1\linewidth]{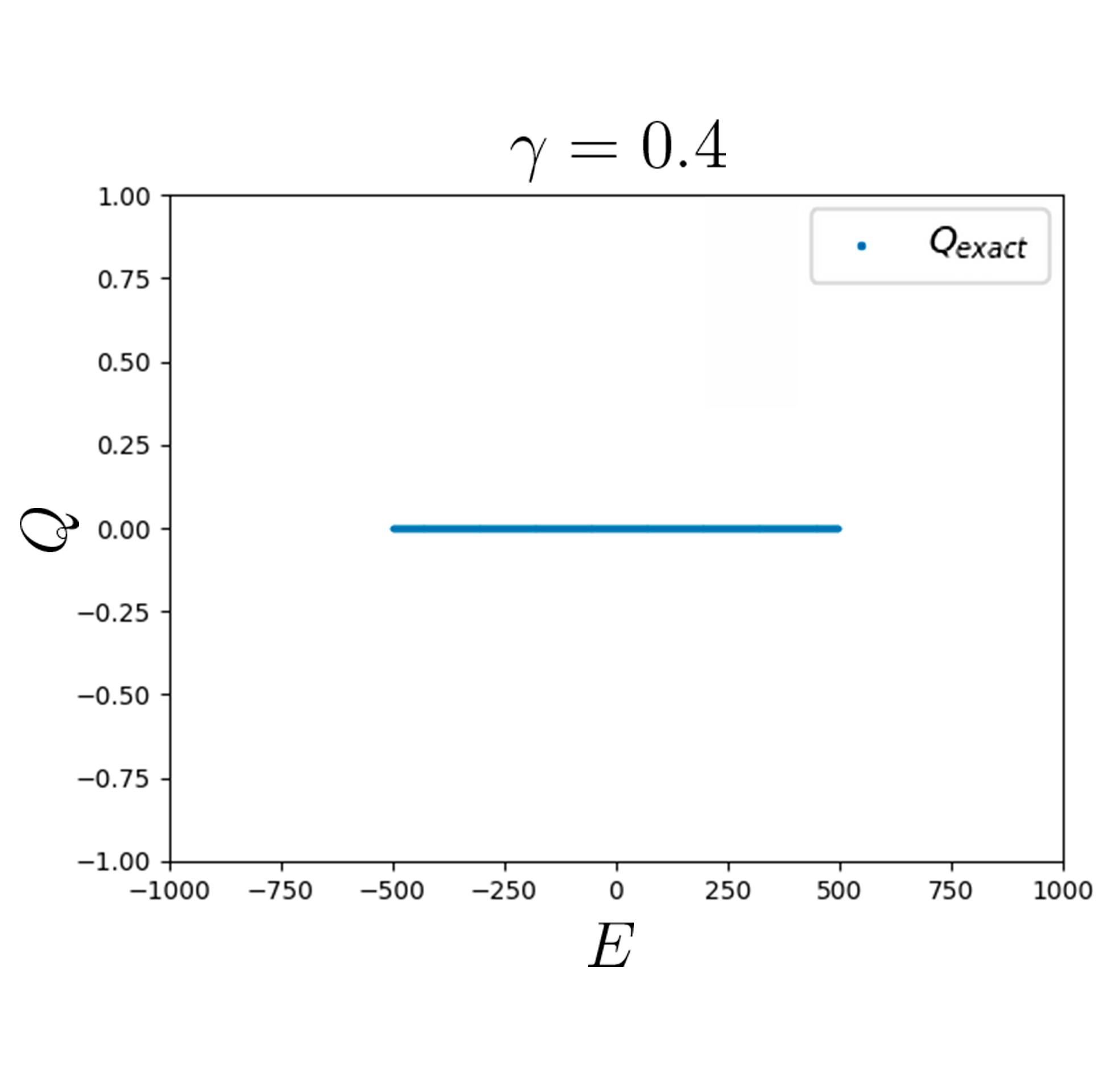} \\(b)}
\end{minipage}
\begin{minipage}[h]{0.325\linewidth}
\center{\includegraphics[width=1\linewidth]{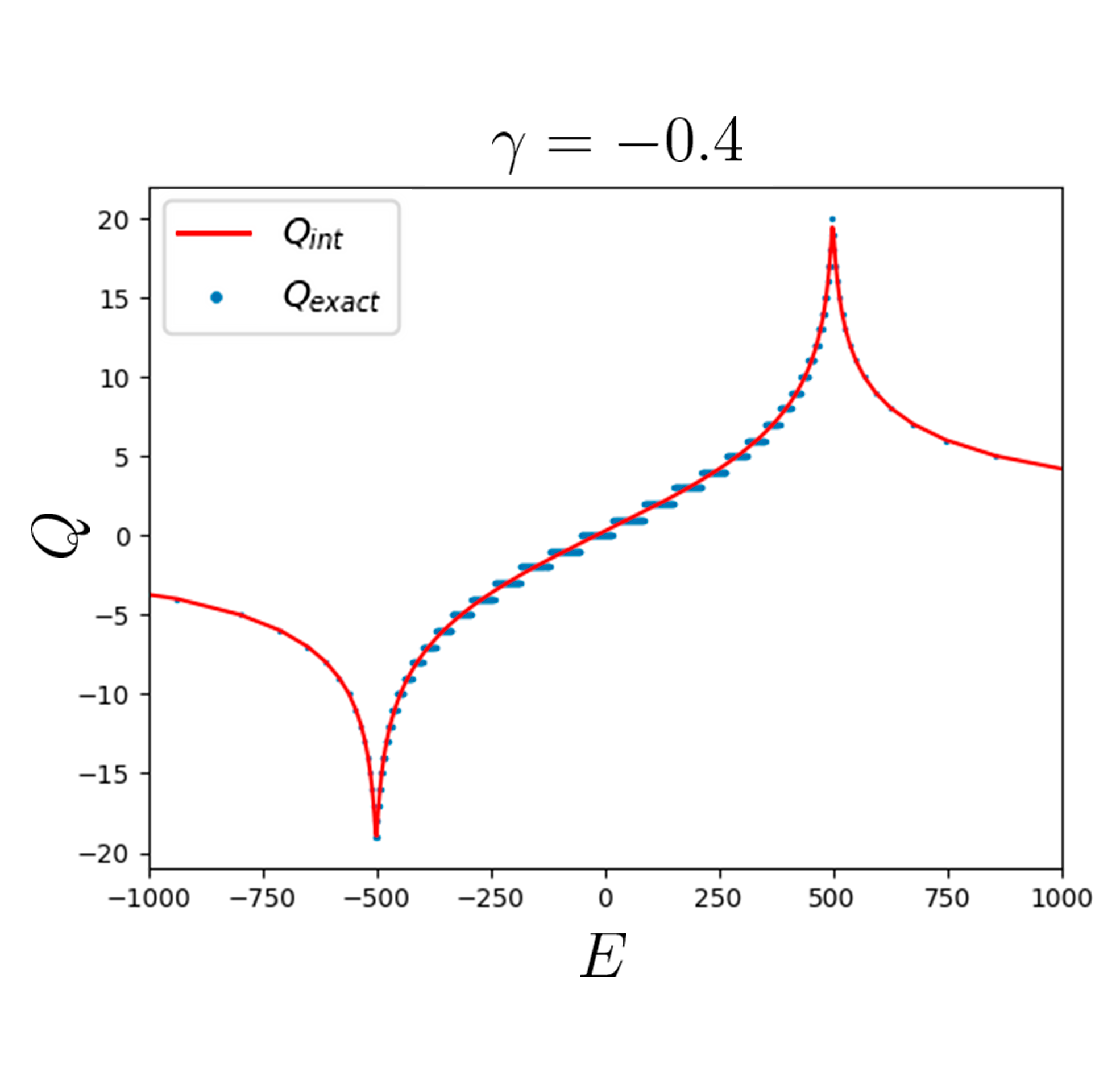} \\(b)}
\end{minipage}
\begin{minipage}[h]{0.325\linewidth}
\center{\includegraphics[width=1\linewidth]{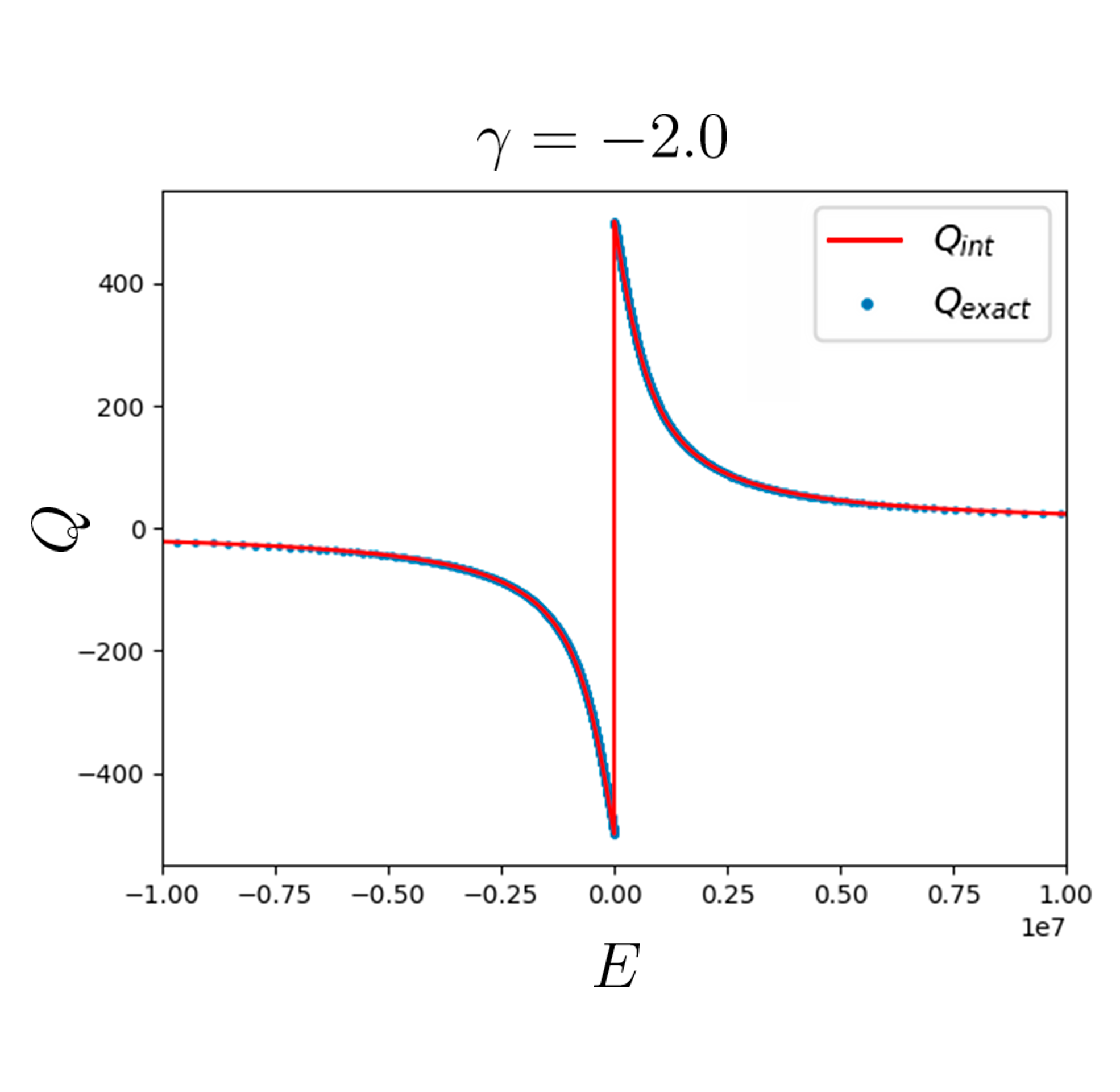} \\(c)}
\end{minipage}
\caption{Distribution of quantum number $Q$ over energies for $N = 1000$, $\theta = \pi/4$, $\delta = 1$ in  (a) localized phase with $Q = 0$. (b) fractal phase with two branches of solutions, levels in the bulk of the spectrum are degenerate with respect to $Q$ (c) delocalized phase with all solution on outer branch. $Q_{int}$ corresponds to Eq.(\ref{BAeq}) where the sum is replaced by an integral, $Q_{exact}$ corresponds to $Q$ obtained for energies from numerical Hamiltonian diagonalization, they are quantized.}
\label{fig:Q(E)}
\end{figure*}
 
\textit{Towers of states}. — The integer $Q$ originates from taking $\ln$ of the Bethe equation and selecting the principal branch of the multivalued $\arctan$ function in Eq.(\ref{BAeq}). 
The equation for $Q$ can be inverted to obtain an approximate expression describing towers of states in the fractal phase:
\begin{equation}\label{towers}\begin{split}
    E^{in}_Q = \frac{\omega}{2} \tanh{\frac{\delta}{2y}(\pi Q - \theta)}\\ E^{out}_Q = \frac{\omega}{2} \coth{\frac{\delta}{2y}(\pi Q - \theta)}
    \end{split}
\end{equation}
There are two different branches of solutions: inner solutions $E_Q^{in}$, located within the interval in $[-N\delta /2, N\delta/2]$ and outer solutions $E_Q^{out}$, located outside this interval. In the localized phase, all solutions are confined near the diagonal elements $\varepsilon_i$ within the interval $[-N\delta /2, N\delta/2]$, however, $Q = 0$ for all levels and no towers of states are formed. As the system progresses to the fractal phase, $Q$ increases, causing the solutions to progressively leave this interval, leading to the formation of two distinct towers characterized by a maximum value $Q_{max}$. In the delocalized phase, the spectrum undergoes a further restructuring: the diagonal elements become irrelevant compared to the off-diagonal part; only the outer tower Eq.(\ref{delocalizedtower}) remains, comprising all solutions. Distribution of $Q$ in different phases is shown in Fig.\ref{fig:Q(E)}.
In the fractal phase $Q_{max}$ and $Q_{min}$ correspond to the levels $E \simeq \varepsilon_{N}$ and $E \simeq \varepsilon_{1}$ respectively.
{
\begin{equation}
    Q_{max} = \frac{y}{\pi\delta}(1 +(\gamma + 1)\ln{N} - \ln{\sin{\theta}} )
\end{equation}}
In the delocalized phase levels with $\pm Q_{max}$ correspond to those with the smallest absolute energy, $\abs{E}$. However, in the delocalized phase, there are two regimes 
{
\begin{equation}\label{Q_max}
    Q_{max}\simeq 
    \begin{cases}
     \beta(N)N\;\;\;\;\;\gamma \in (-2, -1)\\
    \frac{N}{2} \;\;\;\;\;\;\;\;\;\;\;\;\; \gamma < -2
    \end{cases}
\end{equation}}
{with} $\beta(N) \sim O(1)$ \cite{suppl}. The $O(1)$ difference between $Q_{max}$ and $-Q_{min}$ is not significant for the large $N$ limit. We will therefore substitute one for the other, meaning that both determine the height of the tower of states. Although Eq.(\ref{towers}) does not resemble the exponential Efimov scaling of the energies, there are parts of the spectrum that exhibit it. For large  $|Q|$, one can expand Eq.(\ref{towers}) to obtain:
{
\begin{equation}\label{Efimov}
\begin{split}
    E_{Q, Efimov}^{in} \simeq \pm\frac{\omega}{2} \mp \omega e^{-\frac{\delta}{y}(\pi |Q| - \theta)}\\
    E_{Q, Efimov}^{out} \simeq  \pm\frac{\omega}{2} \pm \omega e^{-\frac{\delta}{y}(\pi |Q| - \theta)}\\
\end{split}
\end{equation}}
Roots of this kind are localized near the edges of the diagonal potential, namely $\varepsilon_1$ and $\varepsilon_N$. Moreover, Eq.(\ref{Efimov}) holds at the critical point $\gamma = 0$, where all solutions with non-trivial $Q$ exhibit Efimov scaling.

\textit{RG cycles}. — The renormalization of the Hamiltonian as suggested in \cite{glazek2002limit, Leclair2004russian} is as follows. In each RG step, we take the largest diagonal element $\varepsilon_N$, then reduce the size of the matrix by dropping row and column with this diagonal element, and renormalize $x$ and $y$ to preserve the equations 
for the components of the eigenstates $\psi_i$, 
Eq.(\ref{exact_eig}), $1 \leq i < N$. This leads to the following recurrence relation governing the couplings:
\begin{equation}\label{RG_x}
    x_{N-1} = x_{N} + \frac{x_N^2 + y^2}{\varepsilon_N - E - x_N}
\end{equation}
\begin{equation}\label{RG_y}
    y_{N-1} = y_N = y
\end{equation}
It is convenient to rewrite Eq.(\ref{RG_x}) and Eq.(\ref{RG_y}) in terms of $\theta$ in the same way as was done in \cite{anfossi2005elementary}:
\begin{equation}\label{RG_theta}
    \theta_{N-1} - \theta_{N} = \arctan\frac{y}{E - \varepsilon_N} + \pi Q_N
\end{equation}
Thus, the renormalization group preserves the integrable structure of the model and the Bethe ansatz equations. Since we define $\theta$ to be in $[0, \pi]$, one cycle corresponds to the return of $\theta$ to the initial value with the change of $Q$ by $\pm 1$.
The term containing $\pi Q_N$ can be absorbed into the lhs by introducing a new variable $\tilde{\theta}_N$ with the initial condition $\theta_{N_0}$. Reversing this transformation, $\theta$ can be obtained by taking $\tilde{\theta}$ modulo $\pi$, the number of cycles is determined by $|\theta_N - \tilde{\theta}_N|/\pi$. 

Then, the exact solution to the recurrence relation can be obtained in the form, details are shown in the Supplement \cite{suppl}:
\begin{equation}\label{exactRGtheta}
    \tilde{\theta}_n = \Theta + \Im\ln{\Gamma(n - \frac{N_0}{2} - \frac{E}{\delta} + i\frac{y}{\delta} + 1)}
\end{equation}
where we introduce a parameter independent of $N$: $\Theta(\theta_0, N_0, E, \frac{y}{\delta}) = {\theta}_{N_0}- \Im\ln{\Gamma(\frac{N_0}{2} - \frac{E}{\delta} + i \frac{y}{\delta} + 1) }$. Formula Eq.(\ref{exactRGtheta}) generalizes the result with an energy-dependent RG obtained in \cite{motamarri2024refined}. Since $Q_{min}$ and $Q_{max}$ correspond to $\varepsilon_1$ and $\varepsilon_N$, respectively, we choose the RG transformation so that it eliminates the highest diagonal element $\varepsilon_N$. We can fix the energy at the level $\varepsilon_1$ and observe the evolution of the corresponding $Q_{min}$. In the fractal and localized phases, the solution of recurrence reads as follows: 
\begin{equation}
    \tilde{\theta}_N = \Theta + \frac{y}{\delta} (\ln{N} + O(\frac{1}{N}, \frac{y^2}{\delta^2 N^2})) 
\end{equation}
or in terms of physical coupling $x_N$:
\begin{equation}
    x_N = y \tan{(\frac{y}{\delta}\ln{\frac{N_0}{N}}+\arctan{\frac{x_{N_0}}{y}})}
\end{equation}

In this regime, the RG time is logarithmic, and after one cycle, the size of the system decreases as $N_1 = N_0 e^{-\pi \delta/y}$. In the delocalized phase $\frac{y}{\delta}\gg N$ or $\gamma < -1$:
\begin{equation}
    \tilde{\theta}_N = \Theta + \frac{y}{\delta}(\ln{\frac{y}{\delta}} - 1) + \frac{3\pi}{4} + \frac{\pi N}{2} - \frac{N^2 \delta}{2 y} + O_2
\end{equation}
- where $O_2$ contains terms of order $\frac{N^3}{y^2}$, $\frac{N^4}{y^3}$, etc., that are important for $\gamma > -\frac{3}{2}$, $\gamma > -\frac{4}{3}$, etc. correspondingly. The term $\frac{N^2 \delta}{2 y}$ is important for $\gamma > -2$, and after this point, $N$-dependence is dominated by $\frac{\pi N}{2}$. In this case, $\tilde{\theta}$ depends linearly on the size of the system, and the RG period is $N_0 - N_1 = 2$. The region between logarithmic and linear periodicity, $\gamma \in (-2, -1)$, corresponds to the aperiodic regime, in which $\tilde \theta$ changes by $O(1)$ each step of RG and does not exactly return to its initial value. In the linear periodic regime, Eq.(\ref{RG_x}) can be rewritten as:
\begin{equation}
    x_{N-1} = -\frac{y^2}{x_N}
\end{equation}
After two steps of the RG procedure, $x_N$ returns to its initial value and this matches perfectly with $Q_{min} \simeq -N/2$, as given in Eq.(\ref{delocalizedtower}), which increases by 1 after each RG cycle $N \to N-2$. 
\begin{figure}[h]
    \centering
    \includegraphics[width=1.0\linewidth]{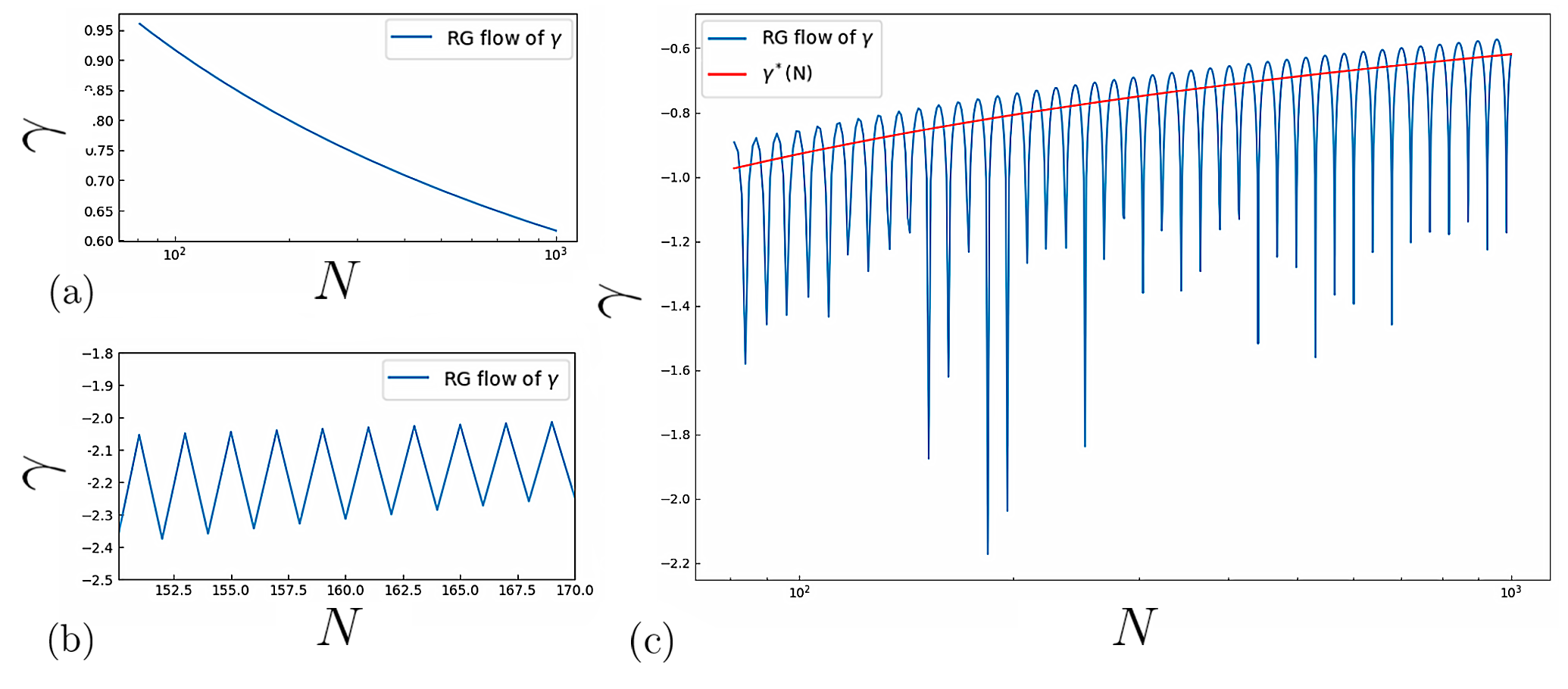}
    \caption{Numerical RG flows of $\gamma$ (a) in the localized phase (b) in the delocalized phase (c) in the fractal phase with flow of $\gamma^{*}$ for initial $\theta_0 = \pi/4$. } 
    \label{gamma_flow}
\end{figure}
The flows of $\gamma$ can be computed from the condition of constant $y$:
\begin{equation}
    \gamma = - \frac{1}{2\ln{N}} \ln{\frac{y^2}{\sin^2{\theta_N}}}
\end{equation}
In this expression, $\gamma$ may take large negative values within the fractal phase. However, this does not contradict our definition of phases, since at these points $\theta$ is close to 0 or $\pi$, keeping $y$ constant. To track the flow of the fractal dimension, it is convenient to monitor $\gamma$ at the points where $\theta$ returns to its initial value with the condition $\frac{\ln{\sin{\theta_0}}}{\ln N}\ll 1$ for the delocalized and fractal phases:
\begin{equation}
    \gamma^{*}(N) = \gamma (N| \;\theta_{N-1} \leq \theta_0 < \theta_{N+1})
\end{equation}

\textit{Order parameter}. — Now we can discuss the RG cycles and the relation of $Q$ to fractality. Let us start the RG algorithm with $N_0$ and $\gamma_0$ in the fractal phase, i.e., the value of $\gamma_0$ in $(-1, 0)$ and $\frac{\ln{\sin{\theta_0}}}{\ln{N_0}}\ll1$, and consider the action of RG in the eigenstate with $E = \varepsilon_1$ and the corresponding quantum number $ \abs{Q_{min}} = Q_{0}$. The value of $\gamma$ after $s$ cycles in the fractal phase is: $\gamma = \gamma_0\frac{\ln{N_0}}{\ln{N}} = \gamma_0\frac{\ln{N_0}}{\ln{N_0} - \pi s\delta/y}$, at cycle $s_1 = \frac{y(\gamma_0 + 1)}{\pi\delta}\ln{N_0}$ the system approaches the delocalized phase. In the delocalized phase, the RG cycle induces a linear change in $N$ both in the aperiodic and linear regimes, as demonstrated in the Supplement \cite{suppl}. Then we can conclude that $(Q_0 - s_1) = \alpha N_0 e^{-\pi s_1 \delta/ y}$, with $\alpha = O(1)$, and as a result we obtain:
\begin{equation}
    \frac{\ln{Q_0}}{\ln{N_0}} = \frac{\ln{y}}{\ln{N_0}} + O(\frac{\ln \ln N_0}{\ln N_0})
\end{equation}
\begin{figure}[h]
\begin{minipage}[h]{0.49\linewidth}
\center{\includegraphics[width=1\linewidth]{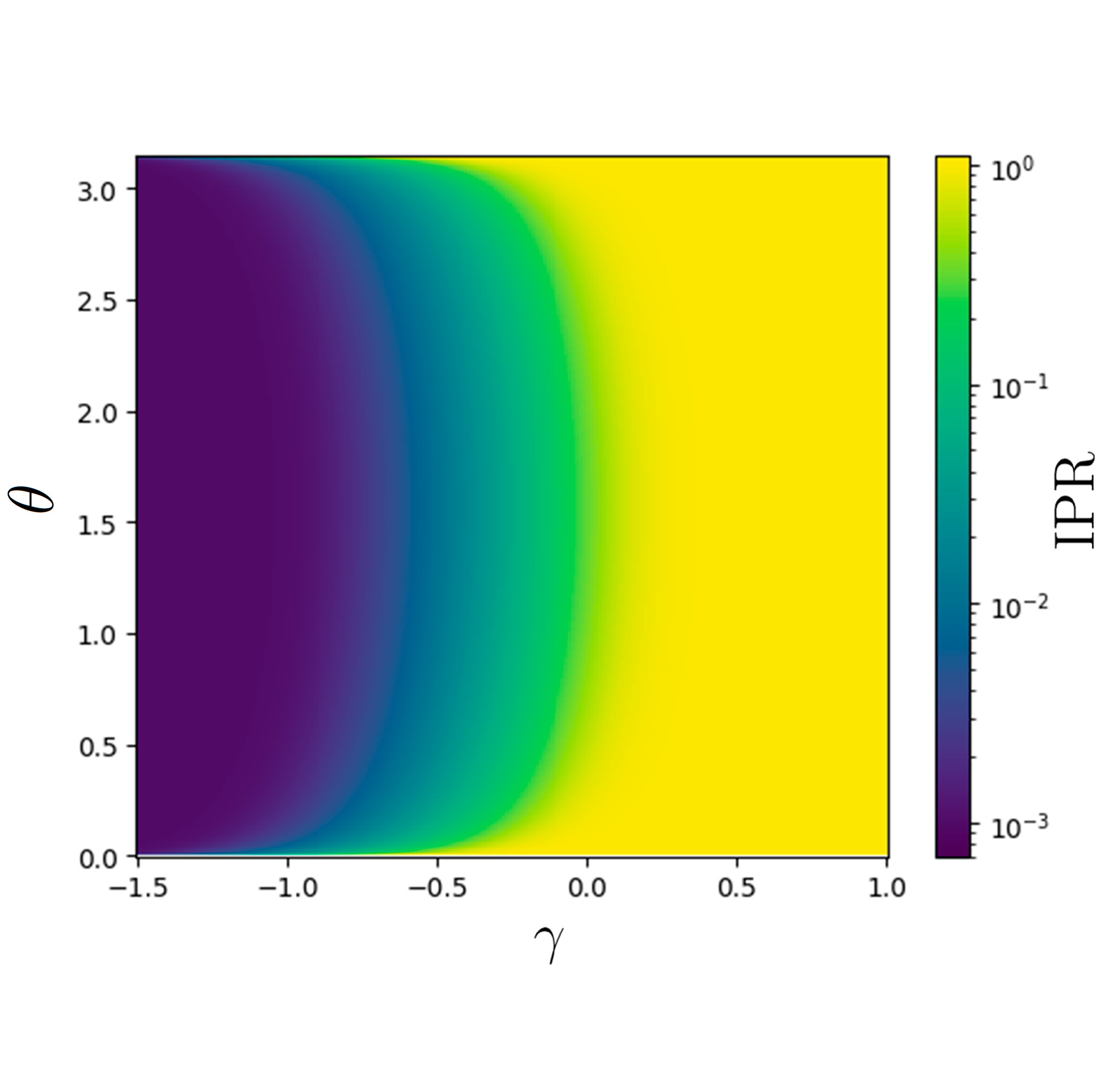} (a)\\ }
\end{minipage}
\hfill
\begin{minipage}[h]{0.49\linewidth}
\center{\includegraphics[width=1\linewidth]{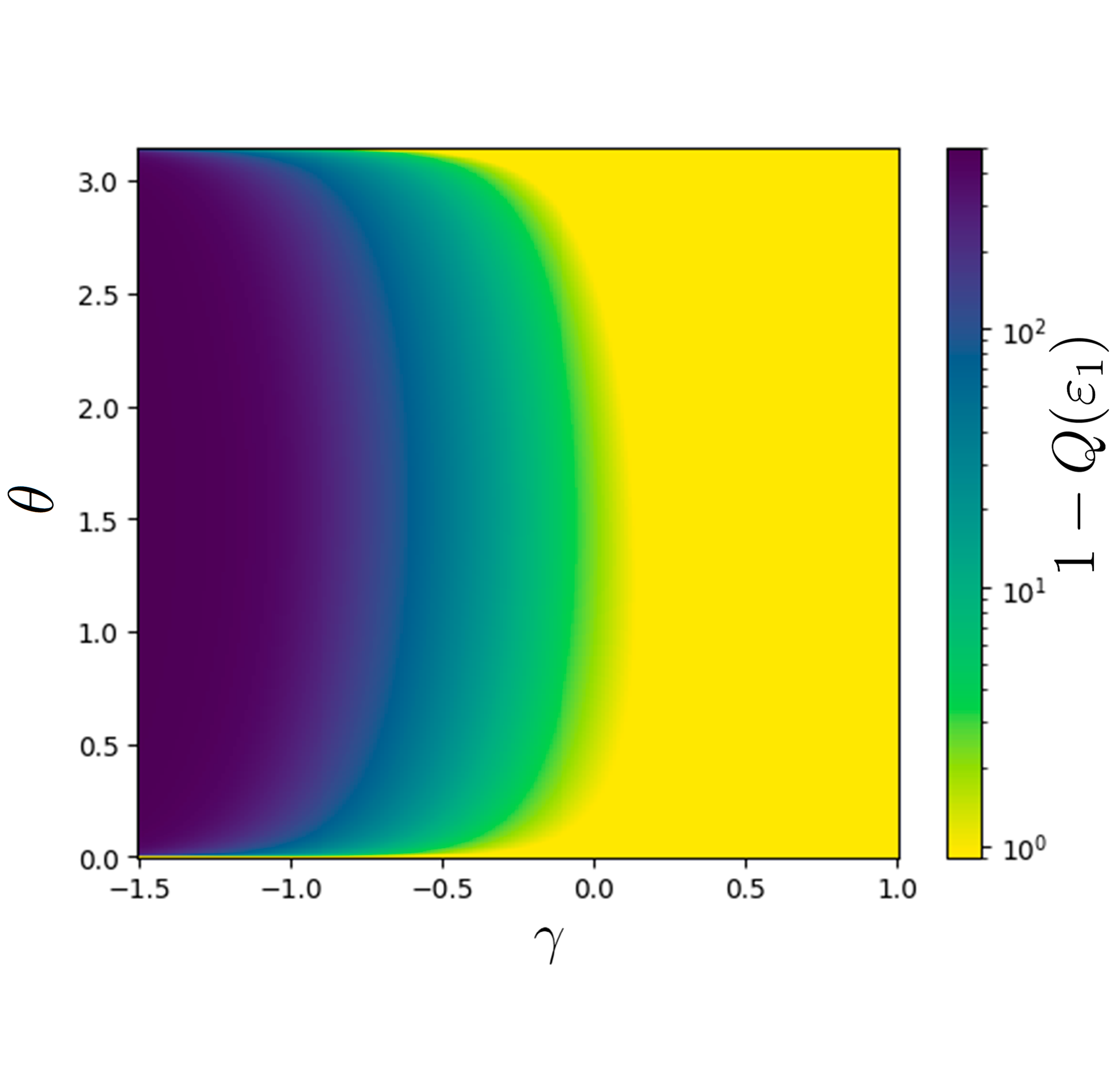} (b)\\}
\end{minipage}
\caption{(a) IPR for $E = \varepsilon_1$ on $(\gamma, \theta)$ plane for $N=1000$, $\delta = 1$. (b) 
{$1 - Q(\varepsilon_1)$} on $(\gamma, \theta)$ plane for $N = 1000$. The 
{addition} of 1 is for convenience of taking $\ln$ in the localized phase.}
\label{IPRQ}
\end{figure}
In the delocalized phase, $Q_{min}$ flows from the value $-\alpha N$ at $\gamma^* = -1$ to $-N/2$ at $\gamma^* = -2$, and then stabilizes at this value in subsequent RG steps. Thus, for the delocalized phase, we obtain:
\begin{equation}
    \frac{\ln{|Q_{min}|}}{\ln N} = 1 + O(\frac{1}{\ln{N}})
\end{equation}
To complete our discussion, we now turn to the localized phase. Upon starting the RG process from the initial value $\theta_0$ with condition $\ln{\sin{\theta_0}}/\ln{N_0}\ll 1$, all degrees of freedom are integrated in less than one RG cycle. This corresponds to $Q_{min} = Q_{max} = 0$. Finally, we have proved the connection between $Q_{min}$ and fractal dimension:
\begin{equation}\label{connection}
    D = \frac{\ln{(1 - Q_{min})}}{\ln{N}} + O(\frac{\ln \ln N}{\ln{N}})
\end{equation}
using renormalization group arguments. $Q_{min}$ is invariant under the shifts of all diagonal elements by a constant. Therefore, Eq.(\ref{Q_max}) for a symmetric distribution remains valid in the process of RG transformations with width $\omega = N \delta$, demonstrating that the RG analysis coincides with the direct computation of the tower of states height. In Fig.\ref{IPRQ} we compare $I_{2} = \mathrm{IPR}$ and the dependence of the order parameter $1 - Q_{min}$ on the $(\gamma, \theta)$ plane.
\begin{figure}[t]
    \centering
    \includegraphics[width=1.0\linewidth]{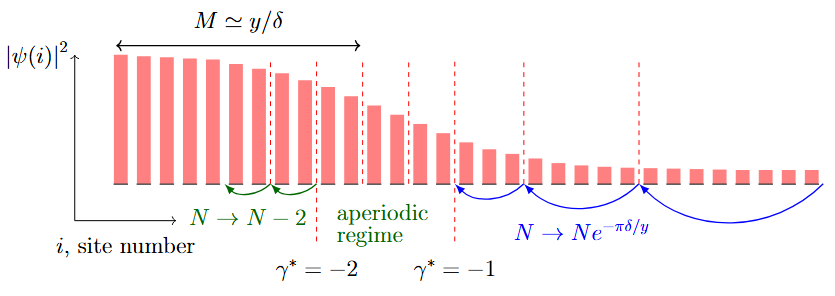}
    \caption{Sketch of RG action on the eigenstate with $E \simeq \varepsilon_1$. $M$ denotes characteristic number of sites, where $\abs{\psi}^2$ is sufficient. RG step corresponds to the elimination of the site with the largest number. Blue arrows represent cycles in the fractal phase with logarithmic RG time, green arrows represent cycles in the delocalized with linear RG time.}
    \label{sketch}
\end{figure}
{Eq.(\ref{connection})} has a clear physical meaning and its sketch is provided in Fig.\ref{sketch}. We start with an eigenstate with $Q = Q_{min}$, which corresponds to the energy level $E = \varepsilon_1$. The probability distribution $|\psi_i|^2$ has a maximum at $i = 1$ and is located on a scale of $1\ll M = \frac{y}{\delta}\ll N_0$ sites. Each RG cycle decreases $|Q_{min}|$ by 1, and the size of the system decreases exponentially in the fractal phase, until it reaches $\gamma^* = -1$ ($N \sim M$). At this point, the system enters the delocalized phase, where the regime becomes aperiodic. This continues until $\gamma^* = -2$ ($M \sim N^2$), at which point periodicity is restored. Although the normalization of the wave function changes during the RG flow due to the elimination of high-energy components, the ratios between the components $|\psi_i|^2/|\psi_j|^2$ are preserved, illustrating the flow towards delocalization.

\begin{figure}[h]
    \centering
    \includegraphics[width=1.0\linewidth]{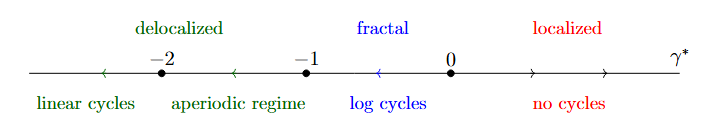}
    \caption{Phase structure of the RD model with flows of $\gamma^*$ } 
    \label{phase_diagram}
\end{figure}
\textit{Geometry of RG flows}. — It is convenient to introduce complex parameter to describe geometry of couplings flows:
\begin{equation}
    z = \theta + i\gamma \ln{N}
\end{equation}
Due to identification of $\theta = \pi$ and $\theta = 0$, the coupling constant manifold is a cylinder $\mathcal{C} = S^1 \times \mathbb{R}$, where $\theta$ is a coordinate on a circle. 
\begin{figure}[h]
    \centering
    \includegraphics[width=1.0\linewidth]{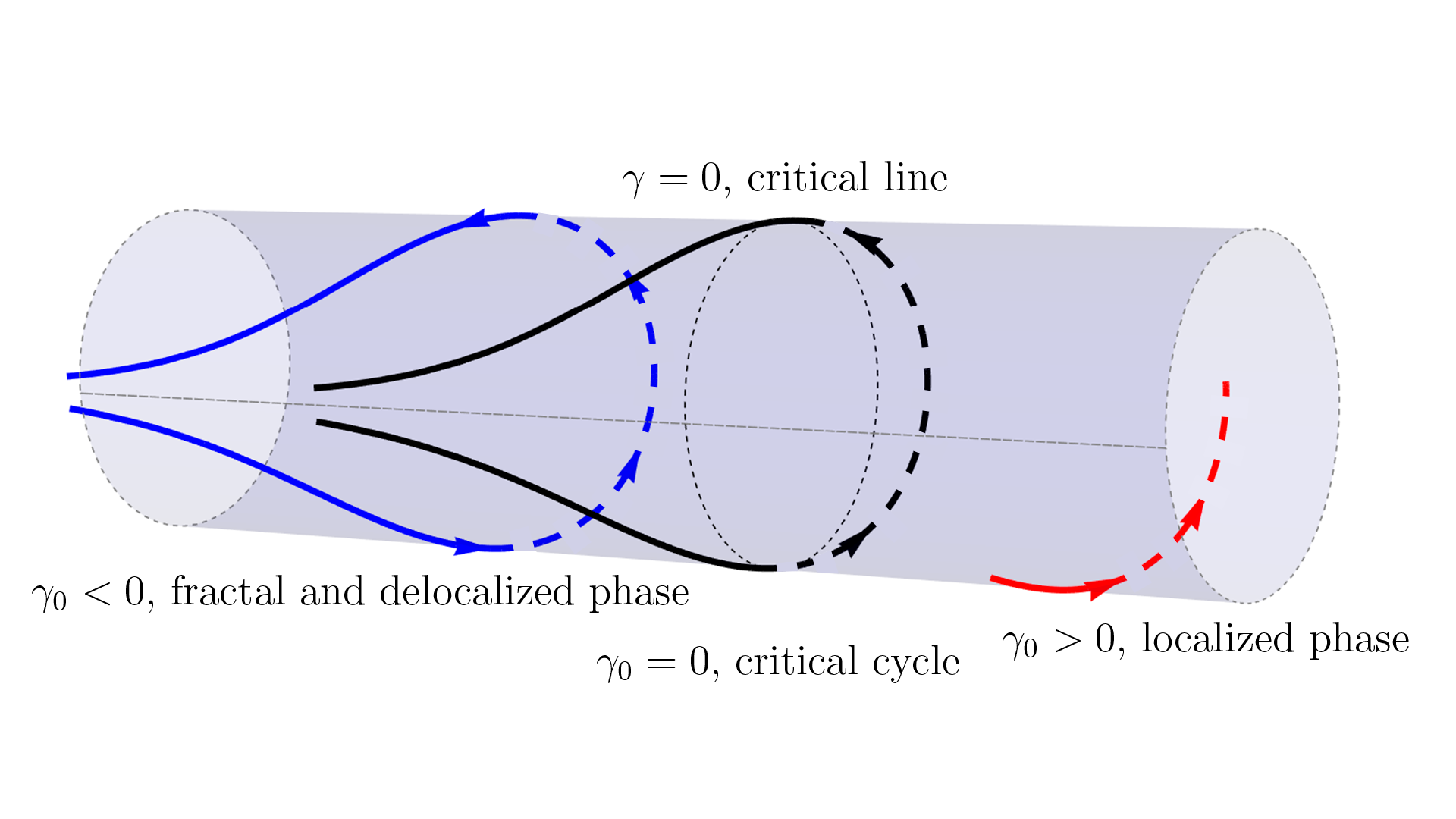}
    \caption{Parameter space manifold $(\gamma\ln N, \theta)$ with examples of flows in different phases.} 
    \label{cyl}
\end{figure}                                                                
Starting from the initial value $\theta_0$ with the condition $\ln{\sin{\theta_0}}/\ln{N} \ll 1$, we obtain different scenarios: for an initial $\gamma_0 > 0$, the process of integrating out of all but one degree of freedom corresponds to an open curve; at $\gamma_0 = 0$, we obtain a critical cycle that intersects with the critical line $\gamma = 0$ after each turn, the integrating out procedure completes after $Q_{max} \sim \ln N$ turns. If we start with $\gamma_0 < 0$ the cycle is the same in both fractal and delocalized phases, the crucial difference lies in the number of RG steps required. In the fractal phase, it takes an extensive number of steps to complete one turn, while in the delocalized phase it takes $\sim O(1)$. In Fig.\ref{cyl} all three types of curve are shown. Consider the map $\phi(\gamma_0, \theta_0): [1, N_0] \to \mathcal{C}$, defined so that $\phi(N) = \theta_N + i\gamma_N\ln N$ and $\phi(N_0) = \theta_0 + i\gamma_0\ln N$. Then, in all phases, $Q_{max}$ is the winding number of the map $\phi$.

{\textit{Discussion.} — The results obtained in this work suggest several directions for future investigation. For the many-body case, an Efimov tower of gaps was obtained in \cite{anfossi2005elementary}, with gaps $\Delta = \Delta_0 e^{-\pi\delta/y}$. However, the spectrum and the properties of the eigenstates in different regions of the parameter space of $(x, y)$ still require a more detailed analysis. In this work, we have shown that the scaling of the maximal number of cycles determines the fractal dimension of the wave functions and, in this sense, acts as an order parameter. Nevertheless, its connection to the breaking of conformal symmetry to the discrete subgroup remains to be clarified. The same applies to the origin of the non-exponential behavior of the state towers discussed in Eq.(\ref{towers}), as well as to the restoration of exponential scaling near the edges of the spectrum. Another promising direction concerns a similar investigation of RG cycles and fractality in the Glazek–Wilson model \cite{glazek2002limit}, which could help establish the universality of the results. Since the RDM is closely related to the twisted inhomogeneous XXX spin chain, the results obtained could provide new insights for the different aspects of the spin chains. 
}

\textit{Conclusion.} — In summary, we have identified the regimes of the renormalization group in the delocalized, fractal, and localized phases of the RD model. In the fractal phase, a cyclic renormalization group is observed with logarithmic RG time, and an Efimov scaling is observed in the particular energy interval. Upon transition to the delocalized phase, the periodicity disappears; however, deeper into the delocalized phase, the periodicity is restored with linear RG time. The mode quantum number in BAE $Q$ simultaneously determines the number of RG cycles and $1 - Q_{min}$ is shown to serve as an order parameter.

The authors thank I. Khaymovich for useful discussions.

\textit{Data availability} — The code reproducing the findings of
this article is openly available at \cite{code}.
\bibliographystyle{unsrt}
\bibliography{ref}

@article{kravtsov2015random,
   title={A random matrix model with localization and ergodic transitions},
   volume={17},
   ISSN={1367-2630},
   url={http://dx.doi.org/10.1088/1367-2630/17/12/122002},
   DOI={10.1088/1367-2630/17/12/122002},
   number={12},
   journal={New Journal of Physics},
   publisher={IOP Publishing},
   author={Kravtsov, V E and Khaymovich, I M and Cuevas, E and Amini, M},
   year={2015},
   month=dec, pages={122002} }

@article{evers2008anderson,
   title={Anderson transitions},
   volume={80},
   ISSN={1539-0756},
   url={http://dx.doi.org/10.1103/RevModPhys.80.1355},
   DOI={10.1103/revmodphys.80.1355},
   number={4},
   journal={Reviews of Modern Physics},
   publisher={American Physical Society (APS)},
   author={Evers, Ferdinand and Mirlin, Alexander D.},
   year={2008},
   month=oct, pages={1355–1417} }

@article{motamarri2022localization,
   title={Localization and fractality in disordered Russian Doll model},
   volume={13},
   ISSN={2542-4653},
   url={http://dx.doi.org/10.21468/SciPostPhys.13.5.117},
   DOI={10.21468/scipostphys.13.5.117},
   number={5},
   journal={SciPost Physics},
   publisher={Stichting SciPost},
   author={Motamarri, Vedant and Gorsky, Alexander S. and Khaymovich, Ivan},
   year={2022},
   month=nov }

@article{Bogomolny2018eigenvalue,
  title = {Eigenfunction distribution for the {Rosenzweig-Porter} model},
  author = {Bogomolny, E. and Sieber, M.},
  journal = {Phys. Rev. E},
  volume = {98},
  issue = {3},
  pages = {032139},
  numpages = {5},
  year = {2018},
  month = {Sep},
  publisher = {American Physical Society},
  doi = {10.1103/PhysRevE.98.032139},
  url = {https://link.aps.org/doi/10.1103/PhysRevE.98.032139}
}

@article{sharipov2026hilbert,
  title={Hilbert space geometry and quantum chaos},
  author={Sharipov, Rustem and Tiutiakina, Anastasiia and Gorsky, Alexander and Gritsev, Vladimir and Polkovnikov, Anatoli},
  journal={Physical Review Research},
  volume={8},
  number={2},
  pages={L022008},
  year={2026},
  publisher={APS}
}

@article{motamarri2024refined,
  title={Refined cyclic renormalization group in Russian doll model},
  author={Motamarri, Vedant and Khaymovich, Ivan M and Gorsky, Alexander S},
  journal={SciPost Physics},
  volume={17},
  number={6},
  pages={157},
  year={2024}
}

@article{Leclair2003russian,
  author={Andr{\'e} {LeClair} and Jos{\'e} Mar{\'i}a Rom{\'a}n and Germ{\'a}n Sierra},
title = {{Russian} doll renormalization group and {Kosterlitz-Thouless} flows},
journal = {Nucl. Phys. B},
volume = {675},
number = {3},
pages = {584-606},
year = {2003},
issn = {0550-3213},
doi = {10.1016/j.nuclphysb.2003.09.032},
url = {https://www.sciencedirect.com/science/article/pii/S0550321303007818},
}

@article{Leclair2004russian,
  title = {Russian doll renormalization group and superconductivity},
  author = {{LeClair}, Andr\'e and Mar\'{\i}a Rom\'an, Jos\'e and Sierra, Germ\'an},
  journal = {Phys. Rev. B},
  volume = {69},
  issue = {2},
  pages = {020505},
  numpages = {4},
  year = {2002},
  month = {Jan},
  publisher = {American Physical Society},
  doi = {10.1103/PhysRevB.69.020505},
  url = {https://link.aps.org/doi/10.1103/PhysRevB.69.020505}
}

@article{Dunning2004integrability,
	title={{Integrability of the Russian doll BCS model}},
	author={C. Dunning and J. Links},
journal = {Nucl. Phys. B},
volume = {702},
number = {3},
pages = {481-494},
year = {2004},
issn = {0550-3213},
doi = {10.1016/j.nuclphysb.2004.09.021},
url = {https://www.sciencedirect.com/science/article/pii/S0550321304007278},
}

@article{Bulycheva2014spectrum,
	author = {K. M. Bulycheva and A. S. Gorskii},
	title = {Limit cycles in renormalization group dynamics},
	publisher = {Physics-Uspekhi},
	year = {2014},
	journal = {Phys. Usp.},
	volume = {57},
	number = {2},
	pages = {171-182},
	url = {https://ufn.ru/en/articles/2014/2/g/},
	doi = {10.3367/UFNe.0184.201402g.0182}
}

@article{Richardson1963restricted,
title = {A restricted class of exact eigenstates of the pairing-force {Hamiltonian}},
journal = {Phys. Lett.},
volume = {3},
number = {6},
pages = {277-279},
year = {1963},
issn = {0031-9163},
doi = {10.1016/0031-9163(63)90259-2},
url = {https://www.sciencedirect.com/science/article/pii/0031916363902592},
author = {R.W. Richardson}
}

@article{Richardson1964exact,
title = {Exact eigenstates of the pairing-force {Hamiltonian}},
journal = {Nuclear Physics},
volume = {52},
pages = {221-238},
year = {1964},
issn = {0029-5582},
doi = {10.1016/0029-5582(64)90687-X},
url = {https://www.sciencedirect.com/science/article/pii/002955826490687X},
author = {R.W. Richardson and N. Sherman},
}

@article{ortiz2005bcs,
  title={BCS-to-BEC crossover from the exact BCS solution},
  author={Ortiz, Gerardo and Dukelsky, Jorge},
  journal={Physical Review A—Atomic, Molecular, and Optical Physics},
  volume={72},
  number={4},
  pages={043611},
  year={2005},
  publisher={APS}
}

@article{anfossi2005elementary,
  title={The elementary excitations of the exactly solvable Russian doll BCS model of superconductivity},
  author={Anfossi, Alberto and LeClair, Andr{\'e} and Sierra, Germ{\'a}n},
  journal={Journal of Statistical Mechanics: Theory and Experiment},
  volume={2005},
  number={05},
  pages={P05011},
  year={2005},
  publisher={IOP Publishing}
}

@article{altshuler1997quasiparticle,
  title={Quasiparticle lifetime in a finite system: A nonperturbative approach},
  author={Altshuler, Boris L and Gefen, Yuval and Kamenev, Alex and Levitov, Leonid S},
  journal={Physical review letters},
  volume={78},
  number={14},
  pages={2803},
  year={1997},
  publisher={APS}
}

@inproceedings{nekrasov2010quantization,
  title={Quantization of integrable systems and four dimensional gauge theories},
  author={Nekrasov, Nikita A and Shatashvili, Samson L},
  booktitle={XVIth International Congress On Mathematical Physics: (With DVD-ROM)},
  pages={265--289},
  year={2010},
  organization={World Scientific}
}

@article{nekrasov2009supersymmetric,
  title={Supersymmetric vacua and Bethe ansatz},
  author={Nekrasov, Nikita A and Shatashvili, Samson L},
  journal={arXiv preprint arXiv:0901.4744},
  year={2009}
}

@article{das2025emergent,
  title={Emergent multifractality in power-law decaying eigenstates},
  author={Das, Adway Kumar and Ghosh, Anandamohan and Khaymovich, Ivan M},
  journal={arXiv preprint arXiv:2501.17242},
  year={2025}
}

@article{leclair2004log,
  title={Log-periodic behavior of finite size effects in field theories with RG limit cycles},
  author={LeClair, Andr{\'e} and Rom{\'a}n, Jos{\'e} Mar{\'\i}a and Sierra, Germ{\'a}n},
  journal={Nuclear Physics B},
  volume={700},
  number={1-3},
  pages={407--435},
  year={2004},
  publisher={Elsevier}
}

@article{dukelsky2004colloquium,
  title={Colloquium: Exactly solvable Richardson-Gaudin models for many-body quantum systems},
  author={Dukelsky, J and Pittel, S and Sierra, G},
  journal={Reviews of modern physics},
  volume={76},
  number={3},
  pages={643--662},
  year={2004},
  publisher={APS}
}

@article{braaten2006universality,
  title={Universality in few-body systems with large scattering length},
  author={Braaten, Eric and Hammer, H-W},
  journal={Physics Reports},
  volume={428},
  number={5-6},
  pages={259--390},
  year={2006},
  publisher={Elsevier}
}

@article{dorey2011quantization,
  title={Quantization of integrable systems and a 2d/4d duality},
  author={Dorey, Nick and Lee, Sungjay and Hollowood, Timothy J},
  journal={Journal of High Energy Physics},
  volume={2011},
  number={10},
  pages={1--42},
  year={2011},
  publisher={Springer}
}

@article{glazek2002limit,
  title={Limit cycles in quantum theories},
  author={G{\l}azek, Stanis{\l}aw D and Wilson, Kenneth G},
  journal={Physical review letters},
  volume={89},
  number={23},
  pages={230401},
  year={2002},
  publisher={APS}
}

@article{ievlev2020string,
  title={String baryon in four-dimensional N= 2 supersymmetric QCD from the 2D-4D correspondence},
  author={Ievlev, E and Shifman, M and Yung, A},
  journal={Physical Review D},
  volume={102},
  number={5},
  pages={054026},
  year={2020},
  publisher={APS}
}

@article{dorey1999bps,
  title={The BPS spectra of gauge theories in two and four dimensions},
  author={Dorey, Nicholas and Hollowood, Timothy J and Tong, David},
  journal={Journal of High Energy Physics},
  volume={1999},
  number={05},
  pages={006},
  year={1999},
  publisher={IOP Publishing}
}

@article{son2001instanton,
  title={Instanton interactions in dense-matter QCD},
  author={Son, DT and Stephanov, Misha A and Zhitnitsky, AR},
  journal={Physics Letters B},
  volume={510},
  number={1-4},
  pages={167--172},
  year={2001},
  publisher={Elsevier}
}

@article{yung2024flowing,
  title={Flowing between string vacua for the critical non-Abelian vortex with a deformation of N= 2 Liouville theory},
  author={Yung, A},
  journal={Physical Review D},
  volume={110},
  number={2},
  pages={025004},
  year={2024},
  publisher={APS}
}

@article{Khaymovich2023fractalset,
  title = {Tuning the phase diagram of a Rosenzweig-Porter model with fractal disorder},
  author = {Sarkar, Madhumita and Ghosh, Roopayan and Khaymovich, Ivan M.},
  journal = {Phys. Rev. B},
  volume = {108},
  issue = {6},
  pages = {L060203},
  numpages = {7},
  year = {2023},
  month = {Aug},
  publisher = {American Physical Society},
  doi = {10.1103/PhysRevB.108.L060203},
  url = {https://link.aps.org/doi/10.1103/PhysRevB.108.L060203}
}

@article{monthus2017multifractality,
  title={Multifractality of eigenstates in the delocalized non-ergodic phase of some random matrix models: Wigner--Weisskopf approach},
  author={Monthus, C{\'e}cile},
  journal={Journal of Physics A: Mathematical and Theoretical},
  volume={50},
  number={29},
  pages={295101},
  year={2017},
  publisher={IOP Publishing}
}

@article{dunning2010exact,
  title={Exact solution of the p+ ip pairing Hamiltonian and a hierarchy of integrable models},
  author={Dunning, Clare and Ibanez, Miguel and Links, Jon and Sierra, Germ{\'a}n and Zhao, Shao-You},
  journal={Journal of Statistical Mechanics: Theory and Experiment},
  volume={2010},
  number={08},
  pages={P08025},
  year={2010}
}

@article{gorsky2025theta,
  title={Theta-term in Russian Doll Model: phase structure, quantum metric and BPS fractality},
  author={Gorsky, Alexander and Liubimov, Ilya},
  journal={Journal of High Energy Physics},
  volume={2026},
  number={5},
  year={2026},
  publisher={Springer}
}

@article{cugliandolo2024multifractal,
  title={Multifractal phase in the weighted adjacency matrices of random Erd{\"o}s-R{\'e}nyi graphs},
  author={Cugliandolo, Leticia F and Schehr, Gr{\'e}gory and Tarzia, Marco and Venturelli, Davide},
  journal={Physical Review B},
  volume={110},
  number={17},
  pages={174202},
  year={2024},
  publisher={APS}
}

@article{khaymovich2020fragile,
  title={Fragile extended phases in the log-normal Rosenzweig-Porter model},
  author={Khaymovich, Ivan M and Kravtsov, VE and Altshuler, BL and Ioffe, LB},
  journal={Physical Review Research},
  volume={2},
  number={4},
  pages={043346},
  year={2020},
  publisher={APS}
}

@article{biroli2021levy,
  title={L{\'e}vy-Rosenzweig-Porter random matrix ensemble},
  author={Biroli, Giulio and Tarzia, Marco},
  journal={Physical Review B},
  volume={103},
  number={10},
  pages={104205},
  year={2021},
  publisher={APS}
}

@article{faoro2019non,
  title={Non-ergodic extended phase of the quantum random energy model},
  author={Faoro, Lara and Feigel’man, Mikhail V and Ioffe, Lev},
  journal={Annals of Physics},
  volume={409},
  pages={167916},
  year={2019},
  publisher={Elsevier}
}

@article{rapp2000high,
  title={High-density QCD and instantons},
  author={Rapp, R and Sch{\"a}fer, Thomas and Shuryak, Edward V and Velkovsky, M},
  journal={Annals of Physics},
  volume={280},
  number={1},
  pages={35--99},
  year={2000},
  publisher={Elsevier}
}

@article{schafer1995chiral,
  title={Chiral phase transition and instanton--anti-instanton molecules},
  author={Sch{\"a}fer, Thomas and Shuryak, Edward V and Verbaarschot, JJM},
  journal={Physical Review D},
  volume={51},
  number={3},
  pages={1267},
  year={1995},
  publisher={APS}
}

@article{sarkar2023tuning,
  title={Tuning the phase diagram of a Rosenzweig-Porter model with fractal disorder},
  author={Sarkar, Madhumita and Ghosh, Roopayan and Khaymovich, Ivan M},
  journal={Physical Review B},
  volume={108},
  number={6},
  pages={L060203},
  year={2023},
  publisher={APS}
}

@article{suppl,
  title={Supplementary material: Exactly Solvable RD Model: RG Cycles Meet
Fractality},
  author={Liubimov, Ilya and Gorsky, Alexander},
  journal={},
  year={2026}
}

@article{witten1979instatons,
  title={Instatons, the quark model, and the 1/N expansion},
  author={Witten, Edward},
  journal={Nuclear Physics B},
  volume={149},
  number={2},
  pages={285--320},
  year={1979},
  publisher={Elsevier}
}

@article{code,
    title={Code repository: Exactly Solvable RD Model: RG Cycles Meet
Fractality},
    author={Liubimov, Ilya and Gorsky, Alexander},
    journal={},
    note = {https://github.com/bionhen/RDM}
}

\newpage
\onecolumngrid
\renewcommand{\theequation}{S\arabic{equation}}
\setcounter{equation}{0}
\newpage
\section{Supplementary material: Exactly Solvable RD Model: RG Cycles Meet Fractality}
\subsection{Eigenstates}
We start by reproducing the expression for the eigenstates as functions of the diagonal elements $\varepsilon_i$ and the parameters $x, y$ in the same way as in \cite{gorsky2025theta}; a similar approach was also applied to obtain the recurrence equation for the eigenstates in \cite{glazek2002limit}. The spectral equation is as follows:
\begin{equation}\label{spectral}
    (H\psi)_i= \sum_{j<i}(-r)e^{i\theta} \psi_j + \sum_{j>i}(-r)e^{-i\theta}\psi_j + (\varepsilon_i -x )\psi_i = E\psi_i
\end{equation}
Subtracting equation $i$ from equation $i+1$ yields the following:
\begin{equation}
    re^{i\theta}\psi_{i} - re^{-i\theta}\psi_{i+1} + (\varepsilon_{i+1} - x)\psi_{i+1} - (\varepsilon_i - x)\psi_i  = E(\psi_{i+1} - \psi_i) 
\end{equation}
\begin{equation}\label{rec}
    \psi_{i+1} = \psi_i \frac{E-\varepsilon_i - i y}{E - \varepsilon_{i+1} + i y}= \psi_i\frac{\rho_{i}}{\rho_{i+1}}e^{-i(\varphi_i+ \varphi_{i+1})}
\end{equation}
Solving this recurrence relation with parametrization $\rho_i = \sqrt{(E -\varepsilon_i)^2 + y^2 }$, $\varphi_i = \arctan{\frac{y}{E - \varepsilon_i}}$, we obtain, after imposing normalization, the following expression for the solution for $l > 1$:
\begin{equation}
    \psi_l = \frac{1}{\sqrt{\sum_{k} \frac{1}{\rho^2_k}}}\frac{e^{-i\varphi_l - i\sum_{j = 2}^{l-1}\varphi_j -i\varphi_1}}{\sqrt{(E -\varepsilon_i)^2 + y^2 }} 
\end{equation}
Particularly interesting is the case $\varepsilon_i = 0$ or alternatively the limit of a large negative $\gamma$, where the diagonal elements become negligible. In this case, the spectrum is indexed by an integer $Q$ in the range $-N/2\leq Q< N/2$:
\begin{equation}
    E_Q  =y\cot \frac{\pi Q-\theta}{N}
\end{equation}
The eigenstates have the form of plane waves with momentum $p_Q = 2\arctan{(r\sin{\theta}/E_Q )}  = 2\frac{\pi Q - \theta}{N}$
\begin{equation}\label{Seq:plain_waves}
\psi_l = \frac{e^{-2i\phi_Q l}}{\sqrt{N}} 
\end{equation}
It is also worth commenting on the connection of this result to \cite{Leclair2004russian}. In the proposed solution, the continuous limit was taken directly in the Schrodinger equation Eq.(\ref{spectral}). However, as we see from the exact solution in Eq.(\ref{exact_eig}), the closest components differ by a phase factor $e^{i\varphi_{i} + i\varphi_{i+1}}$ that does not tend to 1 in the limit $\delta \to 0$. Hence, there is no continuous limit for $\psi$, and therefore an integral equation cannot be formulated.
\newline
{Another comment is on the many-body case. Due to the integrable structure of the model \cite{Dunning2004integrability} the eigenstates in the $M$-pair sector take the form:
\begin{equation}
    B^{\dagger}(E_1)...B^{\dagger}(E_m) \ket{\Omega} 
\end{equation}
with pair creation operator $B^{\dagger}(E) =  \frac{1}{C}\sum_i \frac{1}{E - \varepsilon_i + iy} b^{\dagger}_i$. Here $E_a$ is a Bethe root, in general case these roots can be complex and form Bethe strings on the complex plane \cite{anfossi2005elementary}. Despite the similar structure of the solution, the imaginary part of $E_a$ can effectively change the eigenstate width $\Gamma$. Therefore, results from the one-pair sector provide some intuition, but cannot be straightforwardly applied to the many-body problem.
}
\subsection{Eigenstates fractality}
In this section, we reproduce the derivation of the fractal dimension from \cite{gorsky2025theta}.
We will now use an explicit form of the eigenstates to demonstrate their fractality. 
\begin{equation}\label{Iq_exact}
    I_q = \sum_{i} \frac{1}{(\rho_i \sqrt{C})^{2q}}
\end{equation}
with normalization constant $C = \sum_{i}\frac{1}{\rho^2_i}$. To replace the sum with an integral in Eq.(\ref{Iq_exact}), one requires that $|\psi_i|^2$ varies weakly on the scale of $\delta$. Since $|\psi_i|^2$ has a characteristic scale of $y$, we obtain the condition: $\frac{y}{\delta}\gg 1$ which corresponds to $\gamma < 0$.  
\begin{equation}
    C = \sum_{i}\frac{1}{(E -\varepsilon_i)^2 + y^2} \to \frac{1}{\delta}\int_{-\omega/2}^{\omega/2}d\xi \frac{1}{(E -\xi)^2 + y^2} =\frac{1}{y \delta}(\arctan{\frac{-E + \omega/2}{y}} -\arctan{\frac{-E - \omega/2}{y}})
\end{equation}
\[
    I_q = \frac{1}{C^q}\sum_{i} \frac{1}{((E -\varepsilon_i)^2 + y^2)^q} \to \frac{1}{ C^q \delta}\int_{-\omega/2}^{\omega/2}d\xi \frac{1}{((E -\xi)^2 + y^2)^{q}} = 
\]
\begin{equation}\label{I_q_exact}
   = \frac{1}{ C^q \delta}(\frac{-E + \omega/2}{y^{2q}} {}_2 F_1 (\frac{1}{2}, q,\frac{3}{2}, -\frac{(-E + \omega/2)^2}{y^2}) - \frac{-E - \omega/2}{y^{2q}} {}_2 F_1 (\frac{1}{2}, q,\frac{3}{2}, -\frac{(-E - \omega/2)^2}{y^2}))
\end{equation}
To obtain the fractal dimensions analytically, we will find the asymptotic form of the integral:
\begin{equation}
    \frac{1}{\delta}\int_{-\omega/2}^{\omega/2}d\xi \frac{1}{((E -\xi)^2 + y^2)^q} 
\end{equation}
Changing variables and taking large $N$ for $\gamma \in (-1,0)$: 
\begin{equation}
    \frac{1}{\delta} y^{1-2q}\int_{-(\omega/2  +E )/y}^{(\omega/2-E)/y}d\xi \frac{1}{(\xi^2 + 1)^q} \to \frac{1}{\delta} y^{1-2q}\int_{-\infty}^{\infty}d\xi \frac{1}{(\xi^2 + 1)^q} =  \frac{1}{\delta} y^{1-2q}\sqrt{\pi}\frac{\Gamma(q - 1/2)}{\Gamma(q)}
\end{equation}
The validity of the approximation can be checked by estimating the tail of the integral, $\int_{N^{\gamma}}^{\infty} 1/\xi^{2q}\sim N^{\gamma (1-2q)}$. This term is parametrically smaller than the full integral $\int^{\infty}_{-\infty}$ for $q > 1/2$, which is also an essential condition for the integral's convergence.
\begin{equation}
    I_q = \left(\frac{y}{\delta}\right)^{1-q}\pi^{\frac{1}{2}(1-2q)}\frac{\Gamma(q - 1/2)}{\Gamma(q)}
\end{equation}
Using an explicit expression for $I_q$, we can find the fractal dimension in the region $\gamma \in (-1,0)$:
\begin{equation}\label{D_q_clean}
    D_q =  - \gamma + \frac{\ln{\sin{\theta}}}{\ln{N}} + O(\frac{1}{\ln{N}})
\end{equation}

\subsection{Scaling properties around critical points}
In this section, we discuss critical properties of the RD model.
First, we consider point $\gamma = -1$, with $I_q$ described by Eq.(\ref{I_q_exact}). To understand critical properties, one needs to take the double limit $N\to \infty$, $\gamma \to -1 \pm 0$. One can think of a more general form of limit: $\gamma \to -1 + \epsilon(N)$, where $\epsilon(N) \to \pm 0$. For simplicity, we take $E = -\omega/2$:
\begin{equation}
   I_q (\epsilon, \theta, N) = \frac{ 1}{\arctan^q{\frac{N^{\epsilon}}{\sin{\theta}}}}\frac{N}{N^{q-q\epsilon}\sin^{q} {\theta}} {}_2 F_1 (\frac{1}{2}, q,\frac{3}{2}, -\frac{N^{2\epsilon}}{\sin^2{\theta}}) = N^{1-q} \Phi_{-1}(\epsilon \ln{N} - \ln\sin{\theta})
\end{equation}
The critical behavior of $\epsilon$ is as follows:
\begin{equation}
    \epsilon_{cr} \sim \frac{1}{\ln{N}}
\end{equation}
For $\epsilon$ decreasing faster than $\epsilon_{cr}$ we have:
\begin{equation}
    \lim_{N\to\infty} I_q^{(n)} (\epsilon(N), N) = \lim_{N\to\infty} I_q^{(n)} (-\epsilon(N), N)
\end{equation}
i.e. smooth $I_q$ in the thermodynamic limit. Although for $\epsilon$ decreasing slower than $\epsilon_{cr}$ we have a discontinuity in $D_q$:
\begin{equation}
    \frac{d D_q}{d \gamma}|_{\gamma\to -1 - 0} = 0 \neq \frac{d D_q}{d \gamma}|_{\gamma\to -1 + 0} = -1
\end{equation}
Now, let us turn to the point $\gamma = 0$. For the left limit $\gamma \to 0 - 0$ we have the same scaling as at $\gamma = -1$: 
\begin{equation}
    I_q = \Phi_0 (\epsilon\ln N - \ln \sin\theta)
\end{equation}
However, for the right limit $\gamma \to 0 + 0$ we have another behavior. In the localized phase $E$ turns out to be near $\varepsilon_i$ with corrections of the order of $y$ \cite{gorsky2025theta}, so $I_q$ is independent of $N$ in the leading order: $I_q = \mathrm{const}$.
\subsection{Quantum number $Q$}
We now derive the distribution of $Q$ over energies. Taking the limit of large $N$ we can replace the sum with an integral. This approximation is valid for $y  = \sin{\theta}/N^{\gamma}\gg 1$. This condition holds in both the fractal and delocalized phases ($\gamma < 0$):
\begin{equation}\label{Q_integr}
    Q_{int}  = \frac{\theta}{\pi} + \frac{1}{\pi\delta}\int_{-\omega/2}^{\omega/2} d\xi\arctan{\frac{y}{(E- \xi)}}
\end{equation}
\begin{equation}\label{Q_int}
   Q_{int} =\frac{\theta}{\pi}+ \frac{N}{\pi}\frac{(E  + \frac{\omega}{2})}{\omega} \arctan{\frac{y}{E +\frac{\omega}{2}}}-\frac{N}{\pi}\frac{(E- \frac{\omega}{2})}{\omega} \arctan{\frac{y}{E -\frac{\omega}{2}}}-\frac{yN}{2\pi\omega}\ln{\frac{(E -\omega/2)^2+y^2}{(E +\omega/2)^2+y^2}}
\end{equation}
In the fractal phase $Q_{max}$ and $Q_{min}$ correspond to the levels $E \simeq \varepsilon_{N}$ and $E \simeq \varepsilon_{1}$ respectively.
\begin{equation}
    Q(\varepsilon_{\underset{1}{N}}) \simeq \pm \left( \frac{N}{\pi} \arctan{\frac{y}{\omega}} + \frac{yN}{2\pi\omega}\ln{\frac{\omega^2+y^2}{y^2}} \right)
\end{equation}
In the delocalized phase levels with $\pm Q_{max}$ correspond to those with the smallest absolute energy, $\abs{E}$. However, in the delocalized phase, there are two regimes 
\begin{equation}
    Q_{max}\simeq 
    \begin{cases}
         \frac{y}{\pi\delta}(1 +(\gamma + 1)\ln{N} - \ln{\sin{\theta}})  \;\;\;\;\; \gamma \in (-1, 0) \\
         \beta(N) N\;\;\;\;\;\;\;\;\;\;\;\;\;\;\;\;\;\;\;\;\;\;\;\;\gamma \in (-2, -1)\\
    \frac{N}{2} \;\;\;\;\;\;\;\;\;\;\;\;\;\;\;\;\;\;\;\;\;\;\;\;\;\;\;\;\;\;\;\; \gamma < -2
    \end{cases}
\end{equation}
With parameter:
\begin{equation}
    \beta(N) =  \frac{1}{\pi}\arctan{\frac{y}{\omega}} + \frac{y}{2\pi \omega}\ln({1 + \frac{\omega^2}{y^2}})
\end{equation}
which remain monotonous, $\sim O(1)$ for $\gamma \in (-2,-1)$.
\begin{figure}[h]
    \centering
    \includegraphics[width=0.5\linewidth]{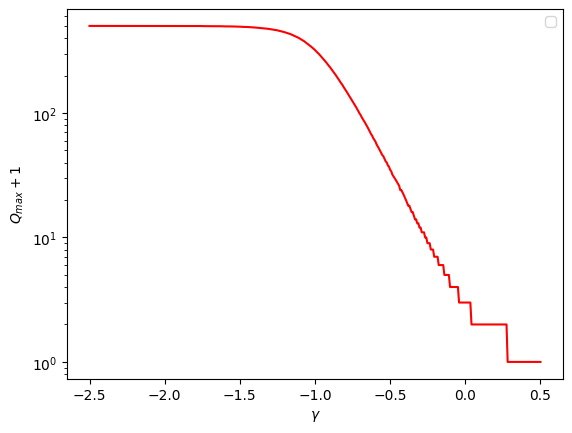}
    \caption{$Q_{max}(\gamma)$ for $\theta = \pi/4$, $\delta = 1$, $N = 1000$ } 
    \label{tower}
\end{figure}
\\One can also reverse equation Eq.(\ref{Q_int}) to obtain the expression for the spectrum in the fractal phase:
\begin{equation}
    E^{in}_Q = \frac{\omega}{2} \tanh{\frac{\delta}{2y}(\pi Q - \theta)}\;\;\;\;E^{out}_Q = \frac{\omega}{2} \coth{\frac{\delta}{2y}(\pi Q - \theta)}
\end{equation}
\subsection{Renormalization group}
Changing the renormalized parameters from $x, y$ to $\theta, y$, one obtains the following renormalization step.
\begin{equation}\label{RG_in_theta}
    \theta_{n-1} - \theta_{n} = \arctan\frac{y}{E - \varepsilon_n} + \pi Q_n
\end{equation}
Here we define $Q_n$ in such a way that $\theta$ stays in $[0, \pi]$ for all RG steps. Once we achieve $\theta_{n_1} = \theta_{n_0}$, $x$ and $y$ return to their initial value, this corresponds to one RG cycle. However, it is not very convenient to solve an equation in this form. Let us introduce $\tilde{\theta}_{n}$ by absorbing $\pi Q_n$. To obtain $\theta_m$ from $\tilde{\theta}_{m}$, one needs to take it modulo $\pi$, $\frac{\abs{\theta_m - \tilde{\theta}_m}}{\pi}$ corresponds to the number of RG cycles. The main step to find the analytical solution is to express the right-hand side as follows: 
\begin{equation}
    \arctan{\frac{y}{E - \varepsilon_n}} = \Im \ln{(1 + i{\frac{y}{E - \varepsilon_n}})}
\end{equation}
with choosing the principal branch of the logarithm. Taking $\varepsilon_i = (i -N/2)\delta$ and using the linearity of $\Im$ and the multiplicativity of the logarithm, we obtain the following.
\begin{equation*}
   \theta_{N_0} - \tilde{\theta}_n = \Im \ln{\frac{\prod_{k = n + 1}^{N_0} (k - \frac{N_0}{2} - \frac{E}{\delta} + i \frac{y}{\delta})}{\prod_{k = n + 1}^{N_0} (k - \frac{N_0}{2} - \frac{E}{\delta})}} = 
\end{equation*}
 where $N_0$ and $\theta_{N_0}$ are initial conditions. Now we can use the property $\Gamma(z+1) = z\Gamma(z)$ of the complex Gamma function to express products as a ratio of Gamma functions:
\begin{equation*}
    =\Im \ln{\frac{\Gamma{(\frac{N_0}{2} - \frac{E}{\delta} + i\frac{y}{\delta} + 1)}}{\Gamma{(\frac{N_0}{2} - \frac{E}{\delta} + 1)}}}\frac{\Gamma{(n - \frac{N_0}{2} - \frac{E}{\delta} + i\frac{y}{\delta} + 1)}}{\Gamma{(n - \frac{N_0}{2} - \frac{E}{\delta} + 1)}}
\end{equation*}
We obtain the following expression, with dependence on $n$ only in the second term:
\begin{equation}
    = \Im\ln{\Gamma(\frac{N_0}{2} - \frac{E}{\delta} + i \frac{y}{\delta} + 1) } - \Im\ln{\Gamma(n - \frac{N_0}{2} - \frac{E}{\delta} + i\frac{y}{\delta} + 1)}
\end{equation}
To explain the connection between fractality and $Q$ we will focus on the energy $E = -N_0/2$, corresponding to the minimal value of $Q$:
\begin{equation}
    \tilde{\theta}_n = \theta_{N_0} - \Im \ln \Gamma(N_0 + i\frac{y}{\delta} + 1) + \Im \ln \Gamma(n + i\frac{y}{\delta} + 1)
\end{equation}
Let us introduce the $n$-independent parameter: $\Theta = \theta_{N_0} - \Im \ln \Gamma(N_0 + i\frac{y}{\delta} + 1)$. This expression can explain the difference between RG in different phases. Since we demand $n \gg 1$ in the RG procedure, we can expand $\ln{\Gamma}$:
\begin{equation}\label{Gamma_asympt}
   \Im \ln \Gamma(n + i\frac{y}{\delta} + 1) = \frac{y}{2\delta}\ln{({(n+1)^2 + \frac{y^2}{\delta^2}})}+   (n + \frac{3}{2})\arctan{\frac{y}{\delta(n+1)}} - \frac{y}{\delta}
\end{equation}
In the fractal and localized phases $\frac{y}{\delta}\ll n$ ($\gamma> -1$):
\begin{equation}
    \tilde{\theta}_n = \Theta + \frac{y}{\delta} (\ln{n} + O(\frac{1}{n}, \frac{y^2}{\delta^2 n^2})) 
\end{equation}
Then $\tilde{\theta}$ depends on the logarithm of the size of the system. The RG period corresponds to the change in $\ln\frac{n_0}{n_1} = \frac{\pi \delta}{y}$. In the delocalized phase $\frac{y}{\delta}\gg n$ ($\gamma < -1$):
\begin{equation}
    \tilde{\theta}_n = \Theta + \frac{y}{\delta}(\ln{\frac{y}{\delta}} - 1) + \frac{3\pi}{4} + \frac{\pi n}{2} - \frac{n^2 \delta}{2 y} + O_2
\end{equation}
- where $O_2$ contains terms of order $\frac{n^3}{y^2}$, $\frac{n^4}{y^3}$ etc., that are important for $\gamma > -\frac{3}{2}$, $\gamma > -\frac{4}{3}$, etc. correspondingly. The term $\frac{n^2 \delta}{2 y}$ is important for $\gamma > -2$, and after this point, $n$-dependence is dominated by $\frac{\pi n}{2}$. In this case, $\tilde{\theta}$ depends linearly on the size of the system, and the RG period is $n_1 - n_0 = 2$. The region between logarithmic and linear periodicity, $\gamma \in (-2, -1)$ (with condition $\ln{\sin{\theta}}/\ln{N} \ll 1$) corresponds to the aperiodic regime.
We defined $Q$ - parameter by choosing the branch of the multivalued $\arctan$ function:
\begin{equation}
    Q(E) =  \frac{\theta}{\pi} + \frac{1}{\pi}    \sum_{l = 1}^N \arctan{\frac{y}{(E - \varepsilon_l)}}
\end{equation}
We choose $E$ to be the closest to the root $\varepsilon_1$ and fix it throughout the RG process. This energy corresponds to the minimal value of $Q$. After each RG cycle, we eliminate high-energy degrees of freedom from Eq.(\ref{RG_in_theta}) 
\begin{equation}
    \tilde{\theta}_{n_2} - \tilde{\theta}_{n_1} = \sum_{i = n_2 + 1}^{n_1} \arctan{\frac{y}{E - \varepsilon_i}} = -\pi
\end{equation}
As a  result, $Q_{min}$ increases by one after every RG cycle. Now we will establish the connection between the RG cycles, $Q$ and the fractality of the eigenstates. We start the RG process in the fractal phase, i.e., the value of $\gamma_0$ in $(-1, 0)$ and $\frac{\ln{\sin{\theta_0}}}{\ln{N}}\ll1$, with $E = \varepsilon_1$ and the corresponding quantum number $ \abs{Q_{min}} = Q_{0}$. Then, the value of $\gamma$ after $s$ cycles is: $\gamma = \gamma_0\frac{\ln{N_0}}{\ln{N}} = \gamma_0\frac{\ln{N_0}}{\ln{N_0} - \pi s\delta/y}$ at cycle $s_1 = \frac{y(\gamma_0 + 1)}{\pi\delta}\ln{N_0}$ the system reaches $\gamma^* = -1$ and transitions to the delocalized phase. In the subsequent aperiodic regime, $\theta$ changes by $O(1)$ at each step and does not return exactly to its initial value. Nevertheless, we can still count the number of cycles that have passed as $|\theta - \tilde\theta|/\pi$. Let us introduce $N_{-1}$ and $N_{-2}$ to denote the system sizes at $\gamma^* = -1$ and $\gamma^* = -2$, respectively. Then, using Eq.(\ref{Gamma_asympt}), the number of cycles between these points is given by: 
\begin{equation}
    \Delta Q = \frac{y}{2\delta}\ln{\frac{\frac{y^2}{\delta^2} + (N_{-2}+1)^2}{\frac{y^2}{\delta^2} + (N_{-1}+1)^2}} + (N_{-2} - N_{-1})\frac{\pi}{2} + (N_{-1}+ \frac{3}{2})\arctan{\frac{\delta (N_{-1} + 1)}{y}} - (N_{-2}+ \frac{3}{2})\arctan{\frac{\delta (N_{-2} + 1)}{y}}
\end{equation}
We also have the condition that $y$ remain constant: $y = N_{-2}^2\sin{\theta_{f}} = N^{-1}\sin{\theta_0}$ with final values $\theta_f$ and $\theta_f + \pi/2$. Beyond the point $\gamma^* = -2$, after many steps, we attain large negative values of $\gamma$, which corresponds to the limit of the plane-wave Eq.(\ref{Seq:plain_waves}) with $Q_{min} \simeq -N/2$, since the period is linear in system size: $\Delta n = 2$, the value of $Q_{min}$ at $\gamma^* = -2$ also behaves as $-N_{-2}/2$. Then $Q_{min}$ at $\gamma^* = -1$:
\begin{equation*}
    Q_{min}(\gamma^* = -1) = -\frac{N_{-2}}{2} - \Delta Q =\frac{N_{-1}\sin{\theta}}{2}\ln{\frac{1}{1+\sin^2 \theta}} - \frac{\pi}{2}N_{-1} + N_{-1}\arctan{\frac{1}{\sin{\theta}}} + O(\sqrt{N_{-1}})
\end{equation*}
It is also convenient to introduce $\alpha$ as follows: $Q_{min}(\gamma^* = -1) = -\alpha N_{-1}$. One can observe that 
\begin{equation}
    \alpha = \frac{1}{2} + \frac{\sin{\theta}}{2\pi}\ln{(1+\sin^2{\theta})} - \frac{1}{\pi}\arctan{\frac{1}{\sin{\theta}}} \leq \frac{1}{2}
\end{equation}
Now we can put all parts together to conclude that $(Q_0 - s_1) = \alpha N_0 e^{-\pi s_1 \delta/ y}$, then ${Q_0}=  \alpha N_0^{-\gamma_0} + \frac{y(\gamma_0 + 1)}{\pi\delta}\ln{N_0}$, as a result, we obtain:
\begin{equation}
    \frac{\ln{Q_0}}{\ln{N_0}} = \frac{\ln{y}}{\ln{N_0}} + O(\frac{\ln \ln N_0}{\ln N_0})
\end{equation}
In the delocalized phase, $Q_{min}$ flows from the value $-\alpha N$ to $-N/2$ through the aperiodic regime. Thus, in the delocalized phase:
\begin{equation}
    \frac{\ln{Q}}{\ln{N}} = 1 + O(\frac{1}{\ln{N}})
\end{equation}
\subsection{On the large $N$ limit}
Remark on the definition of $\gamma$ and the limit $N\to \infty$. Typically, the limit is taken at fixed $\gamma$. In our case, however, we not only work at finite $N$, but also reduce it in the course of the renormalization group procedure with the flow of parameter $\gamma$. Therefore, the question of how to formalize the large limit $N$ arises. 

Consider the initial values in the fractal phase: $-1<\gamma_0< 0$, $\theta_0$ with condition $\ln{\sin{\theta_0}}/\ln{N_0} \ll 1$, then apply the RG transformation with flows: $\theta_N (\theta_0, \gamma_0, N_0)$, $\gamma_N (\theta_0, \gamma_0, N_0)$. Our statement is as follows: in the fractal phase, if we fix $-1<\gamma_1 < \gamma_0$ and define $N_1 = N \;:\; \gamma_N \simeq \gamma_1$ (which defines an implicit dependence $N_1(N_0)$). We use $\gamma_N \simeq \gamma_1$ to denote the condition $\gamma_{N-1} \leq \gamma_1 < \gamma_{N+1}$, since for a discrete RG, the exact value is not attained at finite $N$. Then, if $\ln\sin\theta_{N_1}/\ln{N_1} \to 0$ we obtain the following:
\begin{equation}
   \frac{1}{1- q} \lim_{N_0 \to \infty} \frac{\ln{I_q(\gamma_{N_1}, \theta_{N_1}, N_1)}}{\ln N_1 (N_0)} = \frac{1}{1- q}\lim_{N_1 \to \infty} \frac{\ln{I_q (\gamma_{N_1}, \theta_{N_1}, N_1)}}{\ln N_1} = \gamma_1 
\end{equation}
Taking into account the periodicity of $\theta$ flow, condition $\ln\sin\theta_{N_1}/\ln{N_1} \to 0$ is essentially valid for $\theta_{N_1} \simeq \theta_0$. The flow of $\gamma$ after $p$ cycles yields:
\begin{equation}
    \gamma = \gamma_0\frac{\ln{N_0}}{\ln{N_0} - \pi p N_0^{\gamma_0}}
\end{equation}
$\gamma_1$ is achieved in the cycle: 
\begin{equation}
    p_1 = [(1 - \gamma_0/\gamma_1) \frac{\ln{N_0}}{\pi N_0^{\gamma_0}}] + 1
\end{equation}
- where $[...]$ denotes the integer part of a number. Defining $N_1 = N_0 e^{-\pi p_1 /(N_0^{-\gamma_0}\sin{\theta_0})}$, we can write the following limit:
 \begin{equation}
   \frac{1}{1- q} \lim_{N_0 \to \infty} \frac{\ln{I_q(\gamma_{N_1}, \theta_{N_1}, N_1)}}{\ln N_1 (N_0)} = \frac{1}{1- q}\lim_{N_1 \to \infty} \frac{\ln{I_q (\gamma_{N_1}, \theta_{N_1}, N_1)}}{\ln N_1} = \gamma_1
\end{equation}
In the delocalized phase, if we fix $\gamma_2 < -1$ and define $N_2 = N \;:\; \gamma_N \simeq \gamma_2$, then if $N_2/( N_2^{-\gamma_{N_2} }\sin\theta_{N_2})\to 0$:
\begin{equation}
  \frac{1}{1- q}  \lim_{N_0 \to \infty} \frac{\ln{I_q(\gamma_{N_2}, \theta_{N_2}, N_2)}}{\ln N_2 (N_0)} = \frac{1}{1- q}\lim_{N_2 \to \infty} \frac{\ln{I_q (\gamma_{N_2}, \theta_{N_2}, N_2)}}{\ln N_2 }= 1
\end{equation}
If we start the RG procedure in the fractal phase and achieve a linear periodicity regime in the delocalized phase in the course of the RG transformations, the final values of $\theta$ are $\theta_f$ and $\theta_f + \pi/2$, with $\theta_f$ given by:
\begin{equation}
     \theta_f = \theta_0  + N_0^{-\gamma_0}\sin \theta_0(\ln{(N_0^{-\gamma_0}\sin \theta_0)} -\ln{N_0} - 1) + \frac{3\pi}{4} \;\; \mathrm{mod}\; \pi
\end{equation}
Therefore, one can choose a subsequence $N_{0, k}$ such that $\theta_f(N_{0, k}) = \mathrm{const}$ and take the limit with $\gamma_2 < -2$ and $N_2 = N \;:\; \gamma_N \simeq \gamma_2$ in the following form:
\begin{equation}
    \frac{1}{1- q}\lim_{N_{0, k} \to \infty} \frac{\ln{I_q(\gamma_{N_2}, \theta_{N_2}, N_2)}}{\ln N_2 (N_{0, k})} = \frac{1}{1- q}\lim_{N_2 \to \infty} \frac{\ln{I_q (\gamma_{N_2}, \theta_{N_2}, N_2)}}{\ln N_2 }= 1
\end{equation}
In the localized phase, if we fix the initial values $\gamma_0 > 0$ and $\theta_0$, the flow $\theta_N$ is given by:
\begin{equation}
    \theta_N = \theta_0 + N_0^{-\gamma_0} \sin{\theta_0} \ln{\frac{N}{N_0}}
\end{equation}
Therefore, in the course of RG, the change in $\theta$ is bounded by $|\theta_N - \theta|\leq \delta\theta = N_0^{-\gamma_0} \sin{\theta_0} \ln{N_0}$, with $\delta \theta \to 0$ as $N_0 \to \infty$. The flow of $\gamma$ then can be computed:
\begin{equation}
    \gamma = \gamma_0 \frac{\ln{N_0}}{\ln{N}}
\end{equation}
Thus, if we fix $\gamma_3 > \gamma_0 > 0$, and define $N_3 = N \;:\; \gamma_N \simeq \gamma_3$, we obtain:
\begin{equation}
   \frac{1}{1- q} \lim_{N_{0} \to \infty} \frac{\ln{I_q(\gamma_{N_3}, \theta_{N_3}, N_3)}}{\ln N_3 (N_{0})} = \frac{1}{1- q}\lim_{N_3 \to \infty} \frac{\ln{I_q (\gamma_{N_3}, \theta_{N_3}, N_3)}}{\ln N_3 }= 0
\end{equation}
Since $\delta \theta \to 0$ in the thermodynamic limit, $\theta_N$ remains close to $\theta_0$ for finite $N$, it is convenient to define $\gamma^*_N = \gamma_N$ in the localized phase.
\subsection{Geometry of RG flows}

We have a cylinder geometry due to the identification of $\theta = \pi$ and $\theta = 0$. The flow of $z$ is as follows:
\begin{equation}
    z(N) = \theta_N - i \ln{\frac{y}{|\sin \theta_N|}}
\end{equation}
This expression depends on $N$ only through $\theta_N$, therefore the curve can be parameterized by the values of $\theta$ only and is independent of the RG time. Consider the case of $\gamma_0 = 0$, then $y = \delta\sin{\theta_0}$ and the flow of $z$ is as follows:
\begin{equation}
    \tilde{\theta}_N = \theta_0 + \sin \theta_0 \ln{\frac{N}{N_0}} - i  \ln{\frac{\sin{\theta_0}}{|\sin(\theta_0 + \sin \theta_0 \ln{\frac{N}{N_0}})|}}
\end{equation}
The critical cycle passes through both $\gamma \ln N >0$ and $\gamma\ln N <0$ regions and after each turn $\gamma$ returns to its initial value $\gamma_0 = 0$. In the fractal and delocalized phase we take initial conditions: $\gamma_0 < 0$, $|\gamma_0\log N_0| \gg 1$ and $\ln{\sin{\theta_0}}/\ln N_0 \ll 1$, therefore, there is no intersection with the critical line $\gamma = 0$. This also holds for the localized phase, where initial conditions are as follows: $\gamma_0 > 0$, $\gamma_0\log N_0 \gg 1$ and $\ln{\sin{\theta_0}}/\ln N_0 \ll 1$.
\subsection{ Relation with 4d $N_f=2N_C$ $\cal{N}$=2 SQCD in NS limit}

Comment on the possible application of our findings to 4d $N_f=2N_C$ $\cal{N}$ = 2 SQCD in NS limit. It was found in \cite{gorsky2025theta} that 
the  BA equation for the RDM model with N sites in the M Bethe roots sector exactly coincides with the 
equation for the ground states in the 2d worldvolume theory on the M semilocal
vortex strings \cite{nekrasov2009supersymmetric,nekrasov2010quantization} in the 4d UV finite ${\cal N}=2$ SQCD with $N_F=2N_C$
in the Nekrasov-Shatashvili limit at strong coupling, $\frac{1}{g_{YM}^2}=0,\quad  \theta_{RDM}=\pi -\theta_{4D}$.

The one-pair solution investigated in \cite{gorsky2025theta} corresponds to the single vortex string, and the parameter $Q$ measures the induced electric flux on the vortex worldsheet.
 Electric flux $\langle Q \rangle$ provides the identification of fractal, localized, or delocalized phases 
in the vacuum subsector of the Hilbert space of a vortex string. The very existence of delocalized and fractal phases is a bit unusual in 
2d field theory and requires some comments. The hopping between degenerated  i-th and k-th vacua in the 2d terms means that there
is the kink-antikink pair, hence there is a finite piece of the string in the other vacua, say piece of k-th vacuum in i-th vacuum. Clearly, the kink-antikink
pair annihilates if there is no reason that prevents annihilation. In the non-supersymmetric case the kink-antikink meson state
is legitimate since the vacuum between them is metastable and the vacuum pressure prevents annihilation \cite{witten1979instatons}. In our SUSY case 
the vacua are degenerated, but the flux between kink and antikink changes since the kink acquires the electric charge if $\theta\neq 0$
due to the Witten effect \cite{dorey1999bps}. Therefore, the energy density between them changes and induces the required force. This could explain 
the crucial role of the $\theta$ term in the microscopic explanation of the delocalization and fractality. Note that we consider
the limit $\frac{1}{g_{YM}^2}=0$; therefore, all kinetic terms in 4d theory are suppressed. It would be interesting to link properly
the meaning of the fractality and delocalization  in the vacuum sector of Hilbert space in the 2d worldsheet theory theory and the behavior of the vortex strings in the 4d space-time in the spirit of \cite{altshuler1997quasiparticle}. The relation of the flow between the vacua in the worldsheet theory of the vortex string in this critical regime with the Liouville theory discussed in \cite{yung2024flowing} could be of some use.
Note that there are special massless  baryonic-like states at this 
point in the moduli space \cite{ievlev2020string} that
 could also play a role in the delocalization  of the vortex string in the vacuum subsector.

Our RG procedure corresponds in SQCD to the decoupling of the heaviest flavor 
with mass $\varepsilon_N$ and the RG equation yields the renormalization of the $\theta$-term and the chemical potential. Since 
the masses of  fundamentals and antifundamentals coincide $m_i=\tilde{m}_i$
, two flavors are simultaneously decoupled, and we remain 
in the superconformal point with $N_F\rightarrow N_F-2,\quad N_C\rightarrow N_C-1$. Similarly on the 2d string worldsheet the RG corresponds to the
transition $T^{*}CP(N) \rightarrow T^{*}CP(N-1)$ in the $\sigma$-model target manifold with renormalization of the twist and equivariant parameters. Roughly speaking, the  renormalization comes from the account of the kinks and instantons connecting the remaining vacuum states 
with the single decoupled runaway vacuum.

Let us attempt to  comment on the RG flow in more physical terms.  The $H_{RDM}$ is similar to 
the four-fermion term in dense matter induced by some topological 
configurations, since the hopping terms involve the topological phase $e^{\pm i \theta}$. 
The four-fermion terms in the large $N_C$ limit non-perturbatively can be obtained in two ways.
The single instanton induces a multi-fermionic t'Hooft operator vertex and cannot generate 
a four-fermion term. However, it can be generated via the instanton-anti-instanton pair
configuration \cite{schafer1995chiral} or by fractional instantons 
with topological charges $q=\frac{1}{N_c}$  whose role has been discussed in dense QCD in \cite{son2001instanton, rapp2000high}. Fractional instanton contributions  involve 
a specific factor $e^{i\frac{\theta}{N_C}}$, hence it seems that they cannot 
provide the hopping terms analogous to the RDM Hamiltonian. Therefore we conjecture that the four-fermion vertex is induced by 
the instanton-antiinstanton configuration, however this point certainly deserves a separate detailed study.

\end{document}